\documentclass[aps,twocolumn,superscriptaddress,pre]{revtex4}
\usepackage{graphicx}
\usepackage{dcolumn}
\usepackage{amsmath}
\usepackage{bm}
\usepackage{hyperref}
\usepackage{latexsym}
\usepackage{verbatim}
\usepackage{color}
\usepackage{subfigure}
\usepackage{bm}
\usepackage{float}

\begin{document}

\title{Fluctuations and Pattern Formation in  Self-Propelled Particles}

\author{Shradha Mishra}
\affiliation{Physics Department, Syracuse University, Syracuse, NY
13244}
\author{Aparna Baskaran}
\affiliation{Physics Department, Syracuse University, Syracuse, NY
13244}
\author{M. Cristina Marchetti}
\affiliation{Physics Department, Syracuse University, Syracuse, NY
13244}
\affiliation{Syracuse Biomaterials Institute, Syracuse University, Syracuse, NY
13244}

\date{\today}

\begin{abstract}
{We consider a coarse-grained description of a system of self-propelled
particles given by hydrodynamic equations for the  density and
polarization fields. We find that the ordered moving or flocking state of the system is
unstable to spatial fluctuations beyond a threshold set by the
self-propulsion velocity of the individual units. In this region, the
system organizes itself into an inhomogeneous state of
well-defined propagating stripes of flocking particles interspersed
with low density disordered regions. Further, we find that even in the
regime where the homogeneous flocking state is stable, the system exhibits large fluctuations in both density and orientational order.
We study the hydrodynamic equations analytically and numerically to
characterize both regimes.}
\end{abstract}

\maketitle


\section{Introduction}

Large collections of living organisms exhibit a highly coherent collective dynamics at large scales~ \cite{tonertusr}. This behavior, often referred to as ``flocking",  spans an enormous range of length scales and is seen in diverse systems, including mammalian
herds \cite{parrish}, crowds of pedestrians \cite{helbing1, helbing2}, bird
flocks \cite{starling}, fish schools \cite{hubbard}, insect swarms \cite
{rauch}, bacterial suspensions \cite{benjacob},
extracts of cytoskeletal filaments and molecular motor proteins \cite{harada,
nedelec}, and motility assays \cite{volker}. Many of these systems can be unified under the theoretical paradigm of
collections of self-propelled particles. Their intriguing collective behavior has received considerable attention in recent years.

A number of different theoretical approaches have proved fruitful in understanding the dynamics of collections of self-propelled units. Starting with the seminal work of
Vicsek \cite{vicsek}, rule-based models have been investigated numerically and have been shown to exhibit nonequilibrium transitions between disordered and ordered (flocking or moving) states.
Subsequent work has focused on
characterizing the nature of the order-disorder transition,
its dependence on the noise, and pattern formation
in the ordered state, both in the  context of rule-based Vicsek-type models  \cite{vicsek,chate,aldana} and of models of bacterial swarming~\cite{Peruani2, kaiser1, kaiser2}.
Continuum hydrodynamic theories have been used to describe the behavior of the system at large
scales \cite{tonertuPRL95,RamaswamyPRL,tonertusr,KruseEPJE05}. Self-propelled particles are typically elongated and move along
one direction of their long body axis. They can exhibit orientational order at high concentration. The ordered state is characterized by a vector order parameter, the polarization, which is also proportional to the mean velocity of the system. Hence the ordered state is a macroscopically moving state. The continuum theory has been developed phenomenologically on the basis of general symmetry arguments by drawing on analogies with magnetic systems and with liquid crystals \cite{tonertusr}. In fact active or self-propelled systems  have been likened to ``living liquid crystals"~\cite{Gruler1999}. This work
has yielded several important results, including the possibility of long range order
in 2D \cite{tonertuPRL95} and the prediction and observation of giant number fluctuations in the ordered state\cite{RamaswamyPRL,
RamaswamyEPL}. The continuum theory has also been derived by systematic coarse-graining of specific  microscopic models, including rule-based  \cite{bdg,shradhasr} and physically motivated \cite{aparna,MCM} models. These derivations yields (model-dependent) estimates for the parameters in the hydrodynamic equations and have provided insight into the microscopic origin of of the large-scale collective physics.

\begin{figure*}[tbp]
\begin{center}
\includegraphics[height=12.0cm, width=15.0cm]{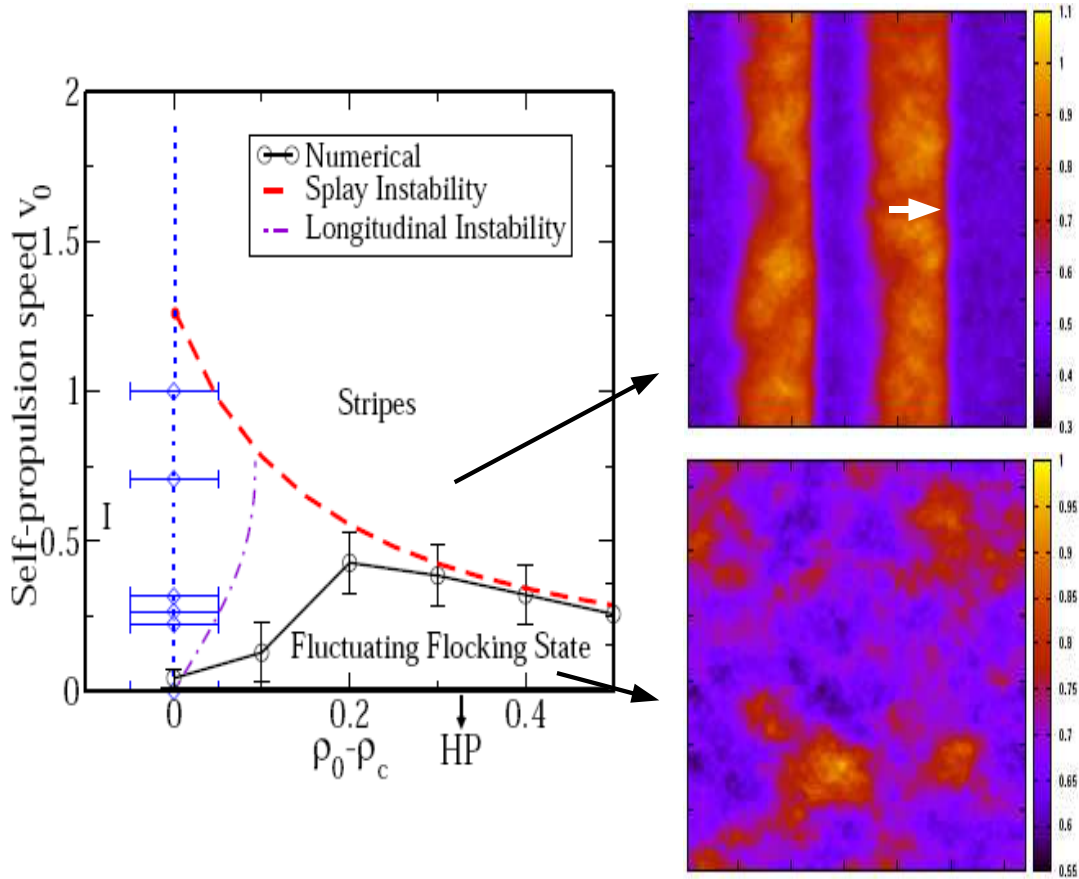}
\end{center}
\caption{(color online) Phase diagram in the $(v_0, \protect\rho_0)$ plane. At $v_0=0$ the system exhibits a continuous mean field transition  at $\rho_0=\rho_c$ from an isotropic (I) to a homogeneous polarized (HP) state. The isotropic phase survives at finite $v_0$ in the region $\rho<\rho_c$ bounded by the vertical dashed line (blue online). For $\rho_0>\rho_c$ there is a critical $v_c(\rho_0)$ separating a polarized moving state with large anomalous fluctuations, named the ``fluctuating flocking state", at low self-propulsion speed from a high-speed phase of traveling stripes. The circles
denote the values of $v_c(\rho_0)$ obtained  numerically with the error bars indicating the step size used in the
computation. The dashed-dotted line (purple online) is the longitudinal instability boundary $v^L_{c1}(\rho_0)$ obtained in Section III.
The dashed line (red online) is the splay instability boundary $v^{S}_c(\rho_0)$, given by Eq.~\eqref{eqvcA}.
The top right
panel is a real space snapshot of the density profile in the striped phase. The stripes travel in the direction of the white arrow, that also denotes the direction of mean polarization in the high density regions. The bottom right panel shows a real
space snapshot of the density profile in the ``coarsening" transient leading to the fluctuating flocking state  at $v_0<v_c$.
Density values from  low (blue) to high (yellow) are indicated in the side bar. The value of the density in the blue stripes is well below the critical value $\rho_c=0.5$, while the red stripes are well into the polarized phase.}
\label{fig:phasedia}
\end{figure*}

In a recent paper we  derived
the hydrodynamic equations for a collections  of self propelled hard rods moving on a frictional substrate and
interacting through excluded volume interactions~\cite{aparna}. Although self-propulsion and steric effects alone are not sufficient to yield a homogeneous polarized moving state in bulk, the hydrodynamic equations are easily modified to incorporate a mean-field continuous transition from an isotropic state at low concentration of rods to a polar state at high density. In the present paper we examine
analytically and numerically the coupled nonlinear hydrodynamic equations for density and polarization to characterize the large-scale structures that replace the linearly unstable homogeneous ordered state.  The main results of our work are summarized
in  Fig.~\ref{fig:phasedia} that represents a  ``phase diagram" in terms of the density $\rho$ of rods and their self-propulsion speed, $v_0$. In the absence of self-propulsion ($v_0=0$) the model considered exhibits a mean-field continuous transition at the critical density $\rho_c$ from an isotropic state of zero polarization for $\rho_0<\rho_c$ to an ordered moving state, with uniform density and macroscopic polarization ${\bf P}\not=0$. In a uniform ordered state at finite $v_0$, all rods would move with uniform mean velocity $\sim v_0 {\bf P}$. We find however that the moving state for $\rho_0>\rho_c$ exhibits more complex behavior. For $v_0$ below a critical value $v_c(\rho_0)$ the steady state of the system is still macroscopically polarized on average, but exhibits anomalous density and polarization fluctuations. We refer to this state as the ``fluctuating flocking state". The anomalous density fluctuations  are  the giant number fluctuations predicted  by Toner and coworkers \cite{tonertuPRE98} and observed experimentally in active nematics \cite{rodsexpt}. An additional feature of this regime,  is a very slow temporal approach to this noisy steady state,  with  some features of a coarsening process. For $v_0>v_c(\rho_0)$ the system orders in a robust striped phase, consisting of traveling  high/low density stripes. The high density stripes are ordered with polarization transverse to the long direction of the stripes, which is also the direction of the stripes' motion.
%
Both of these phases have been identified in numerical studies of the
Vicsek model \cite{chate,onemore?}. Here we characterize them in terms of
their origins in the hydrodynamic equations using both analytical and
numerical tools. This allows us to derive an understanding that transcends
any specific microscopic model and is generically applicable to a large
number of self-propelled systems.

The coupling of density and polarization fluctuations embodied in the convective terms in the hydrodynamic equations of self-propelled systems plays a crucial role in controlling pattern formation. Some of these convective terms reflect the dual role played by the polarization field as vector order parameter and as the mean velocity, resulting in  competition between diffusion and convection along the direction of mean local order. There is a qualitative analogy here with sedimentation problems \cite{sedimentation_simul,sedimentation_theory}, where the interplay of local alignment along the sedimenting direction and diffusion can  destabilize the system resulting in convective patterns, although hydrodynamic interactions, not incorporated here, often also play an important role in sedimenting systems.

The layout of the paper is as follows. First we introduce  the hydrodynamic
equations that are the starting point of our analysis. Then, we carry out a
linear stability analysis about the ordered state and characterize the
region of linear stability of the bulk ordered phase or homogeneous flock. Next,
we report the results of numerical solution of the nonlinear hydrodynamic  equations
and identify and characterize both the fluctuating flocking state and the striped phase, as well as the coarsening-like behavior leading to these phases.

\section{The model}

We consider a collection of polar rods of length $\ell $
moving on an inert substrate characterized by a friction constant $\zeta $ in two dimensions (2d).
Each rod is driven by an internal  force $F$ acting along one direction of its long axis,
called its head. This force, together with the frictional interaction with
the medium, results in a
self-propulsion speed  $v_{0}=F/\zeta$ of constant magnitude. On length scales long compared to $\ell $
and on time scales long compared to the microscopic interaction times, the
dynamics of the system can be described  in terms of hydrodynamic fields,
namely the conserved densities
(here the density $\rho\left( \mathbf{r},t\right) $ of rods) and the variables associated with possible
broken symmetries.  A collection of self-propelled polar rod can order in a polarized state, characterized by a finite value of a vector order parameter,   ${\bf P}({\bf r},t)$, describing the mean polarization of the rods. The ordered state is also a moving state, with mean velocity $\sim v_0{\bf P}$. The dynamics of the system is described by coupled equations for density and polarization, given by
\begin{equation}
\partial _{t}\rho =-\bm\nabla \cdot \left( \rho v_{0}\mathbf{P}-D\bm\nabla\rho \right)  \label{nonlinDens}
\end{equation}
and
\begin{widetext}
\begin{eqnarray}
\partial _{t}\rho \mathbf{P}+\lambda _{1}\rho \mathbf{P}\cdot \bm\nabla \rho
\mathbf{P} &=&-D_{r}\left[ a_{2}\left( \rho \right) +P^{2}a_{4}\left( \rho
\right) \right] \rho \mathbf{P}-\frac{v_{0}}{2}\bm\nabla \rho +\lambda _{3}\rho P_{i}%
\bm\nabla \rho P_{i}+\lambda _{2}\rho \mathbf{P}\bm\nabla \cdot \rho \mathbf{%
P}  \notag \\
&&+\left( D_{s}-D_{b}\right) \bm\nabla \left( \bm\nabla \cdot \rho \mathbf{P}%
\right) +D_{b}\nabla ^{2}\rho \mathbf{P}.  \label{nonlinPol}
\end{eqnarray}

\end{widetext}
{
 The hydrodynamic  equations \eqref{nonlinDens} and \eqref{nonlinPol} have the same form as those first proposed on a phenomenological basis by Toner and Tu~
\cite{tonertuPRL95,tonertuPRE98,tonertusr} to describe the physics of flocking. The  parameter  $a_2$ is chosen to change sign at a characteristic density $\rho_c$, while $a_4>0$.  This guarantees a mean-field continuous transition from an isotropic state with $\rho=\rho_0$ and ${\bf P}$=0 when $a_2>0$ to a homogeneous polarized state with $\rho=\rho_0$ and $|{\bf P}_{0}|=\sqrt{\frac{
-a_{2} }{a_{4}}}$ when $a_2<0$. These equations have also been derived from specific microscopic models of self-propelled particles on a substrate  by some of us~\cite{ABMCMPRE2008,aparna,ABMCMJStPh2010} and by Bertin and collaborators~\cite{bdg}.
Bertin et al obtained hydrodynamic equations by coarse-graining a Vicsek-type model of self-propelled {\em point particles}, with a specific aligning rule for the pair interaction. In contrast, Baskaran and Marchetti, considered a model of self-propelled hard rods of finite size with excluded volume interaction and analyzed in detail the modifications induced by self-propulsion on the linear and angular momentum exchanged in a binary collision. The equations obtained by Bertin et have precisely the form given in Eqs.~\eqref{nonlinDens} and \eqref{nonlinPol}, with parameters $a_2$ and $a_4$ determined by the aligning interaction between particles. In contrast, it was demonstrated in Ref.~\cite{aparna} that steric effects alone are not sufficient to yield a homogeneous bulk polarized state. As a result, the equations derived in \cite{aparna} for a purely physical model have $a_4=0$. We note that it was also demonstrated recently in a model of swimmers in a fluid that hydrodynamic interactions among swimmers are equally insufficient to yield a homogeneous polarized state in bulk \cite{ABMCM_PNAS}. These results  suggest that genetically and biochemically-regulated
signaling, or external symmetry-breaking effects, such as chemotaxis, may be needed to obtain a polar state. Non-physical or external mechanisms of this type  are embodied in Bertin et al in an alignment rule, but are absent in the work by Baskaran and Marchetti that aimed at identifying the role of purely physical interactions in controlling the large scale behavior of self-propelled systems. The goal of the present paper is to study the stability of the homogeneous polar state. For this reason we have added phenomenologically the term proportional to $a_4$ to the equations derived in \cite{aparna}. Both $a_2$ and $a_4$ will then treated as phenomenological parameters.

 The density satisfies a conservation law, with a flux controlled by two terms:  $\rho v_{0}\mathbf{P}$ describing convection along the mean self-propulsion velocity, $v_0{\bf P}$,  and a diffusive current $-D\bm\nabla\rho $
that drives the system to a homogeneous state. The anisotropy of the diffusion coefficient relative to the direction of mean motion is neglected here for simplicity.   The various terms  in Eq.~(\ref{nonlinPol}) (other than the one proportional to $a_4$) are obtained from the microscopic hard rod model and have a simple physical interpretation. The polarization field ${\bf P}$ plays a dual role in self-propelled systems. On one hand, it represents the vector order parameter associated with the spontaneous
breaking of rotational symmetry in the polarized state.
Its dynamics is then in the class of that of equilibrium polar liquid
crystals and X-Y spin systems.  On the other hand, $v_{0}
\mathbf{P}$ is also the mean velocity of the flock with which particles are convected. The interplay between these two physical roles of the polarization field gives rise to the various terms in Eq.~(\ref
{nonlinPol}) and underlies most of the phenomena discussed in this paper.
The three terms proportional to $\lambda_i$ in Eq.~(\ref
{nonlinPol}) play a crucial role in controlling the pattern
formation phenomena described below. If we think of $v_0{\bf P}$ as a velocity, then all three terms have the structure of convective nonlinearities. Galilean invariance would  require $\lambda _{1}=\frac{v_{0}}{\rho }$ and $\lambda_2=\lambda_3=0$.
The self-propelled overdamped system considered here is, however,  moving relative to a fixed substrate and does not satisfy Galilean invariance. As a result, the values of $\lambda _{i}$ are unconstrained and
in general model-dependent. There is an additional important difference between these three terms. The term proportional to $\lambda_1$ is a truly nonequilibrium term that can be understood only as a convective nonlinearity. In contrast, the terms proportional to $\lambda_2$ and $\lambda_3$ have an equilibrium-like interpretation associated with the role of  $\mathbf{P}$ as the polar order parameter. These two terms would arise in equilibrium from a term of the form
$\rho\left| \mathbf{P}\right| ^{2}\bm\nabla \cdot \mathbf{P}$ in the free energy, which effectively accounts for a dependence of the elastic constant associated with splay deformations on the amount of orientational order in the system. In this case one would obtain $\lambda_2=-\lambda_3$.
Finally, $D_{s}$ and $D_{b}$ are diffusion constants that characterize the
relaxation of splay and bend fluctuations, respectively, and $D_{r}$ is a rotational diffusion rate.

Before proceeding to analyze the hydrodynamic equations, it is useful to introduce dimensionless variables.
 We measure
time in units of the inverse rotational diffusion rate, $D_{r}^{-1}$, and lengths in units of the length $\ell$ of the self propelled particles. The various dimensionless fields and parameters are then given by
\begin{equation*}
\tilde{\rho}= \rho \ell ^{2},
\end{equation*}
\begin{equation*}
\tilde{v_{0}}= v_{0}/(\ell D_{r}),
\end{equation*}
\begin{equation*}
\tilde{\lambda} _{i}= \lambda _{i}/(D_{r}\ell ^{3}),
\end{equation*}
\begin{equation*}
\tilde{D}= D/(D_{r}\ell ^{2}).
\end{equation*}
In the following all quantities are dimensionless and we drop the tilde for simplicity of notation.


\section{Linear Stability}
The hydrodynamic Eqs.~(\ref{nonlinDens}) and (\ref{nonlinPol}) admit two homogeneous solutions: an isotropic state (I) with $\rho=\rho_0$ and ${\bf P}=0$ for $\rho<\rho_c$, and a homogeneous polarized (HP) state with $\rho=\rho_0$ and ${\bf P}\equiv P_0\hat{\bf p}_0$ for $\rho_0>\rho_c$, where $\mathbf{\hat{p}}_{0}$ is the direction of broken symmetry and
$P_0=\sqrt{-a_2/a_4}$. The critical value $\rho_c$ is defined by $a_2(\rho_c)=0$ and is chosen here as $\rho_c=0.5$. It was shown in Ref.~\cite{aparna} that the isotropic state is always linearly stable. In this section we examine the linear stability of the HP state at finite $v_0$.
To do this we linearize the hydrodynamic equations by letting
\begin{eqnarray}
&&\rho =\rho _{0}+\delta \rho, \\
&&\mathbf{P}=\mathbf{\hat{p}}_{0}(P_0+\delta P)+P_{0}\delta \mathbf{p}_{\bot }\;,
\end{eqnarray}
where $\mathbf{\hat{p}}_{0}\cdot\delta \mathbf{p}_{\bot }=0$.
Inserting this ansatz in Eqs.~(\ref{nonlinDens}) and (\ref{nonlinPol}), we obtain
three coupled equations for the fluctuations in the density, $\delta\rho$, the
magnitude $\delta P$ of the polar order parameter and the director, $\delta{\bf p}_{\bot }$.  Combining the fluctuations into a vector,
\begin{equation}
\delta y_{\alpha }\left( {\bf r},t\right) \rightarrow \left(
\begin{array}{c}
\delta \rho \left( {\bf r},t\right)/\rho_0  \\
\delta P\left( {\bf r},t\right)  \\
\delta {\bf p}_{\bot }\left( {\bf r},t\right)
\end{array}
\right)
\end{equation}
and introducing the Fourier components, $\delta \widetilde{y}\left( \mathbf{k},t\right) =\int_{\mathbf{r}}e^{i\mathbf{k }\cdot \mathbf{r}}\delta y\left(
\mathbf{r},t\right)$, the coupled linear equations can be written in matrix form as
\begin{equation}
\delta \widetilde{y}_{\alpha }\left( {\bf k},t\right) =A_{\alpha \beta
}\left( {\bf k} \right) \delta \widetilde{y}_{\beta }\left( {\bf k}%
,t\right)\;,\label{eqnFormal}
\end{equation}
where
\begin{widetext}
\begin{equation}
{\bf A}\left( {\bf k}\right) =\left(
\begin{array}{ccc}
ik_\Vert v_{0}P_{0} -Dk^{2} & ik_\Vert v_{0}  & ik_\perp v_{0}
P_{0}  \\
\begin{array}{c}
-2\alpha a_{20} P_{0}+ik_\Vert\left(
\frac{v_{0}}{2}+\overline{\lambda }\rho^2 _{0}P_{0}^{2}\right)
  \\
-D_{s}k_\Vert^{2} -D_{b}k_\perp^2
\end{array}
&
\begin{array}{c}

2a_{20} -ik_\Vert\overline{\lambda }\rho _{0}P_{0}  \\
-D_{b}k_\perp^2-D_{s}k_\Vert^2
\end{array}
&
\begin{array}{c}
-ik_\perp\lambda _{2}\rho _{0}P_{0}^{2}
-( D_{s}-D_{b})k_\Vert k_\perp
\end{array}
\\
ik_\perp\left( \frac{v_{0}}{2P_{0}}-\rho _{0}P_{0}\lambda _{3}\right)
-\left( D_{s}-D_{b}\right) k_\Vert k_\perp
  &
\begin{array}{c}
-ik_\perp \lambda _{3}\rho _{0}
-\frac{\left( D_{s}-D_{b}\right) }{P_{0}}k_\Vert k_\perp
\end{array}
&
\begin{array}{c}
ik_\Vert\lambda _{1}\rho _{0}P_{0}
-D_{s}k_\perp^2-D_{b}k_\Vert^2
\end{array}
\end{array}
\right) \label{fullLinear}
\end{equation}
\end{widetext}
with ${\bf k}={\bf \hat{p}}_0k_\Vert+{\bf k}_\perp$, ${\bf k}_\perp=k_\perp{\bf \hat{k}}_\perp$, $\delta\tilde{y}_3({\bf k},t)={\bf \hat{k}}_\perp\cdot\delta{\bf \tilde{p}}_\perp({\bf k},t)$,
 and
\begin{equation}
\alpha =\frac{\rho_{0}}{2a_{20}}\bigg(\frac{\partial a_2}{\partial \rho}\bigg)_{\rho = \rho_0}-\frac{\rho_{0}}{2a_{40}}\bigg(\frac{\partial a_4}{\partial \rho}\bigg)_{\rho = \rho_0},
\label{alpha}
\end{equation}
where $a_{20} = a_2(\rho_0) <0$, $a_{40} = a_4(\rho_0) >0$,
and
\begin{equation}
\overline{\lambda}=\lambda _{1}-\lambda _{2}-\lambda _{3}.
\label{lambdabar}
\end{equation}
The coefficients $a_2$ and $a_4$ are  chosen of the simplest form that guarantees a continuous transition at $\rho_c$ and $P_0\simeq 1$ for $\rho_0\gg\rho_c$, i.e.,
\begin{subequations}
\begin{gather}
a_2=1-\rho/\rho_{c}\;,\\
a_4=1+\rho/\rho_{c}\;.
\label{a2a4}
\end{gather}
\end{subequations}
With this choice $\alpha=\frac{\rho_{0}\rho_{c}}{\rho^{2}_{0}-\rho^2_{c}}$  is always positive.

%
%
We look for solutions of the form
$\delta\widetilde{y}({\bf k},t)\sim e^{s_\alpha\left( \mathbf{k}\right) t}$, where the rates $s_\alpha\left( \mathbf{k}\right) $ are the hydrodynamic modes of the system. These are defined as those with decay rates (here proportional to $Re[s_\alpha(k)]$) that vanishes in the long wavelength limit $k\rightarrow 0$. Modes with $Re[s_\alpha(k)]<0$ decay at long times, while modes with $Re[s_\alpha(k)]>0$ grow, rendering  the homogeneous state linearly unstable.
We discuss the hydrodynamic modes by considering some simplified  cases. Further details are given in Appendix A.

First, we consider  the behavior of the system for $\rho_0\gg\rho_c$, i.e.,  deep in the ordered state. The rate of decay of long wavelength fluctuations of  the magnitude $\delta P$ of the order parameter is controlled by $A_{22}\sim 2 a_{20}$, which is always finite for $\rho\gg\rho_c$, away from the mean field continuous transition. In other words $\delta P$ is a nonhydrodynamic variable that decays on microscopic time scales. In this regime we can then neglect fluctuations  $\delta P$ and simply consider the  dynamics of density and director fluctuations
governed by the two coupled equations
\begin{widetext}
\begin{equation}
\partial _{t}\delta \widetilde{\rho }=\left( ik_\Vert v_{0}P_{0}
-Dk^{2}\right) \delta \widetilde{\rho }+ik_\perp v_{0}\rho _{0}P_{0}
\delta \widetilde{p}_{\bot }  \label{dens.deep}
\end{equation}
\begin{equation}
\partial _{t}\delta \widetilde{p}_{\bot } =\left[i\left( \frac{v_{0}}{
2P_{0}}-\rho _{0}P_{0}\lambda _{3}\right) -k_\Vert \left( D_{s}-D_{b}\right)
 \right] k_\perp \frac{\delta \widetilde{\rho }}{\rho _{0}}
+\left[ ik_\Vert \lambda _{1}\rho _{0}P_{0} -\left( D_{s}k_{\perp}
^{2} +D_{b}k_{\Vert}^{2} \right) \right] \delta \widetilde{p}_{\bot }
\label{pol.deep}
\end{equation}
\end{widetext}
The general form of the dispersion relation of the hydrodynamic modes is readily obtained by solving a quadratic equation and is given in Appendix A. Here we discuss some limiting cases.
For wavevectors ${\bf k}$ along the direction $\hat{\bf p}_0$ of broken symmetry, i.e., $k=k_\Vert$ and $k_\perp=0$, density and orientation fluctuations decouple and decay with rates
\begin{subequations}
\begin{gather}
s^L_\rho(k)=ikv_0P_0-Dk^2\;,\label{longrho}\\
s^L_p=ik\lambda_1\rho_0P_0-D_bk^2\;.
\label{longp}
\end{gather}
\end{subequations}
Both modes are stable and propagating, albeit with different speeds. Deep in the ordered region, $P_{0}\simeq 1$.
The propagation speed of density fluctuations is then simply $ v_{0}$,
 while the propagation speed of director
fluctuations  is $\lambda _{1}\rho _{0}\sim v_0^2 \rho_0$. A
Galilean invariant system would have  $\lambda _{1}=v_{0}/\rho_0$ and the two modes would have the same propagation speed, $v_0$. The difference in the propagation  speed
of the two modes can then be considered a signature of the violation of Galilean invariance.

Next, we consider wavevectors ${\bf k}$ transverse to the direction $\hat{\bf p}_0$ of broken symmetry, i.e., $k=k_\perp$ and $k_\Vert=0$. In this case the equations for density and director fluctuations are coupled and the two hydrodynamic modes are given by
\begin{widetext}
\begin{equation}
s_{\pm }^{T}=-\frac{1}{2}\left( D+D_{s}\right) k^{2}\pm
\frac{1}{2} \Big\{\left( D-D_{s}\right)
^{2}k^{4}-2k^{2}v_{0}\rho _{0}\left[
v_{0}-2\rho _{0}P^{2}_{0}\lambda _{3}\right] \Big\}^{1/2}\;.
\label{splaymodes}
\end{equation}
\end{widetext}
The mode $s_{+}^{T}$ can become positive, yielding an instability, for $k<k_c$,
with
\begin{equation}
k_c=\sqrt{v_0\left[2\rho_0P_{0}^{2}\lambda_3-v_0\right]/(D D_s)}\;,
\end{equation}
%
%
provided
\begin{equation}
2\rho_0P_{0}\lambda_3>v_0\;.
\label{inst}
\end{equation}
The parameter $\lambda_3$ has been estimated for a few microscopic models and found to be of order $v_0^2$ \cite{bdg,aparna}.  Eq.~(\ref{inst}) then identifies a value $v^{S}_c(\rho_0)$ of the self propulsion speed above which the homogeneous polarized state becomes unstable, with $k_c\sim( v_0-v^S_c)^{1/2}$. The instability boundary  $v^{S}_c(\rho_0)$  depends on microscopic parameters and is therefore model-dependent.
Using the parameter values obtained for the model of self-propelled hard rods discussed in \cite{aparna},
where the nonlinear terms in the polarization equation arise from
momentum-conserving collisions between the self-propelled rods,
and summarized in Table~\ref{table1}, we obtain
\begin{equation}
v^{S}_c=\left[2\pi\rho _{0}P_{0}^{2}\right]^{-1}.
\label{eqvc}
\end{equation}
%
This instability is associated with splay deformations of the director and  with spatial gradients  normal to the direction of mean order, suggesting that it may bear no relevance to the stripe formation observed in the numerics due to the fact that the particles in the stripes are always aligned along the short direction thereby retaining the long-wavelength nature of the splay mode. However, as shown in Fig.~(\ref{fig:phasedia}), the splay instability line agrees remarkably well with the numerically observed onset of stripe formation at high density. This result, discussed further below, suggests that nonlinear pattern selection mechanisms  may play an  important role in stripe formation.
Finally,  Eq.~\eqref{eqvc} shows that $v^{S}_{c}\sim \frac{1}{\rho _{0}}$ for $\rho_0>>\rho_c$ and appears to diverge as we approach the phase transition. This apparent divergence is regularized when the effect of overdamped fluctuations of the magnitude of the polar order is incorporated in the mode analysis. This is done in Appendix A, where rather than just neglecting $\delta\tilde{P}$ entirely, we approximate its behavior by assuming that on the time scales of interest $|\partial_t\delta\tilde{ P}|\ll |2a_{20}\delta\tilde{P}|$ in  Eqs.~\eqref{eqnFormal} and \eqref{fullLinear}. We then  neglect the time derivative of $\delta\tilde{ P}$, solve for the overdamped $\delta\tilde{P}$,  and use this result to eliminate it from the equations for  density and director fluctuations. We find that the resulting modes still exhibit a splay instability as described above, but the critical speed monotonically approaches the constant value  $v^{S}_c(\rho_c) = [\frac{\pi}{2} \rho_c]^{-1}$ as $\rho _{0}\rightarrow\rho _{c}$. The renormalized boundary $v^{S}_c(\rho_0)$ of this splay instability given in Eq.(\ref{eqvcA}) is plotted in Fig.~\ref{fig:phasedia} as a red dashed line. In addition,
fluctuations in the magnitude of the polarization renormalize of the diffusion constant associated with the decay of density fluctuations.

As the continuous order-disorder transition is approached from above, $a_{20}\rightarrow 0$ and the separation of time scales between the decay of speed/magnitude fluctuations $\delta P$ and the true hydrodynamic variables $\delta \rho$ and $\delta p_{\bot}$ no longer holds. To capture the physics of the system in the vicinity of the order disorder transition, we need to retain the dynamics of the "non-hydrodynamic" variable $\delta P$ and examine the three coupled equations (\ref{eqnFormal}). One can show that the splay instability described above for ${\bf k}$ normal to the direction of mean order survives and is qualitatively unchanged. On the other hand, for ${\bf k}$ along the direction of broken symmetry,  director fluctuations $\delta p_{\bot}$ decouple from density and speed fluctuations and decay at the rate \eqref{longp}.
 The coupled modes for the dynamics of density and speed fluctuations are then given by
\begin{equation}
s_{\pm }^{L}=\frac{1}{2}\left( A_{11}+A_{22}\right) \pm \frac{1}{2}\sqrt{%
\left( A_{11}-A_{22}\right) ^{2}+4A_{21}A_{12}}\label{modesformal2}
\end{equation}
where $A_{ij}$ are the elements of the matrix ${\bf A}({\bf k},t)$ given in Eq.~(\ref{fullLinear}) for ${\bf k}=k_\Vert{\bf \hat{p}}_0$.
It is easy to see that one of the two dispersion relations describes a non-hydrodynamic mode,  $s_{+}^{L}\left( 0\right) =2 a_{20}$, but with a decay rate that becomes vanishingly small for $\rho_0\rightarrow\rho_c$. The other mode vanishes at  $ k=0$.
At small wavevectors  the dispersion relation of the hydrodynamic mode $s^{L}_{-}$ takes the form
\begin{equation}
s_{-}^{L}\left( k\right) = ikv_0P_0(\alpha+1)-D_{eff}k^{2}+{\cal O}(k^3)\;,
\end{equation}
with
\begin{equation}
D_{eff}=D+\frac{v_{0}^{2}}{4\left| a_{20}\right| }-\frac{
v_{0}^{2}\left( \alpha +1\right) ^{2}}{2a_{40}}+\frac{\pi v_{0}^{3}\rho
_{0}\left( \alpha +1\right) }{2a_{40}}     \label{deff}
\end{equation}
where we have used the microscopic parameters given in Table 1. When $D_{eff}<0$,  density fluctuations  grow in time and the ordered state is unstable. At the phase transition, i.e., for $\rho_0=\rho_c$, The condition $D_{eff}<0$ is satisfied for all values of $v_0$
and the ordered state is always unstable. Away from the transition, by considering the exact modes in (\ref{modesformal2}) we find that for densities in a range $\rho_c\leq \rho_0 \leq \rho^{L}_{c}$ there exists a range of self propulsion speeds $v^{L}_{c1}\leq v_{0}\leq v^{L}_{c2}$ where the propagating density fluctuations are unstable. The lower  instability boundary $v^{L}_{c1}(\rho_0)$ is shown in Fig.~(\ref{fig:phasedia}) as a purple dashed-dotted line.

\begin{figure}[tbp]
\begin{center}
{\includegraphics[height=9.0cm, width=9.0cm]{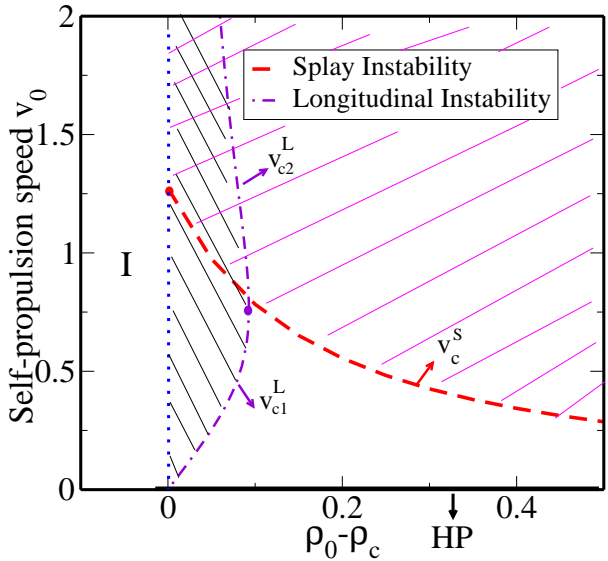}}
\end{center}
\caption{(color online) The figure displays the linear stability boundaries in the $(v_0,
\rho_0)$ plane. All lines have been calculated using the microscopic parameter values of Table \ref{table1}.  The vertical dotted line (blue online) is the mean field continuous transition from the isotropic (I) to the homogeneous polarized (HP) state. The dashed-dotted lines (purple online) are the boundaries (calculated by numerical solution of   $D_{eff}(v_0,\rho_0)=0$, with $D_{eff}$ given by Eq.~\eqref{deff})  that define the region $v_{c1}^L\leq v_0\leq v_{c2}^L$ where the homogeneous polarized state is unstable due to the growth of coupled density and polarization fluctuation associated with spatial gradients along the direction of mean order (longitudinal instability). The linear theory predicts that the homogeneous polar state is unstable in the ruled region to the right of the vertical mean-field transition and bounded by these two lines.  The dashed line (red online) is the splay instability boundary given in Eq.~\eqref{eqvcA}. It terminates at a finite value at $\rho_0=\rho_c$. The linear theory predicts that splay fluctuations destabilize the polar state in the entire ruled region above the dashed (red) line. The region where the system exhibits both the longitudinal and splay instabilities is cross-hatched. The  longitudinal instability boundary $v_{c1}^L(\rho_0)$ vanishes for $v_0\rightarrow 0$ and $\rho_0\rightarrow\rho_c^+$, in agreement with the numerics.}
\label{fig:linear-stab}

\end{figure}

The results of the linear stability analysis are summarized in Fig.~\ref{fig:linear-stab}. The linear stability analysis predicts that near the mean field order-disorder transition the homogeneous ordered state is destabilized at small $v_0$ by the growth of coupled density and polarization fluctuations. The instability occurs for spatial gradients along the direction of mean order, signaling the onset of spatial structures that are inhomogeneous in this direction, like the stripes found numerically. The  wavevector $k_c$ of the fastest growing mode for this instability scales as $k_c\sim (v_0-v_c)^{-1/2}$ at fixed density $\rho_0$ and as $(\rho_0-\rho_c)^{1/2}$ at fixed self propulsion speed.   The boundary of stability $v^{L}_{c1}(\rho_0)$   obtained from the linear theory vanishes at $\rho_c$, in agreement with the onset of the striped phase  obtained by numerical solution of the nonlinear equations,  as shown in Fig.~(\ref{fig:phasedia}), but grows faster  with $v_0$ than obtained numerically.  This discrepancy is likely to stem from the fact that the {\em full nonlinear dynamics} of amplitude fluctuations must be incorporated to account for the behavior in these regions. In addition, the linear stability analysis predicts that the homegeneous ordered state is again stable at large self-propulsion speed for $v_0>v^{L}_{c2}(\rho_0)$. This second line is shown in Fig.~\ref{fig:linear-stab}. The numerics, however, yield a striped phase in this region. Finally,  the longitudinal instability only exists for  $\rho_0\leq\rho^{L}_{c}\simeq 1.1$, while numerically stripes are observed at all densities above a critical velocity. Deep in the ordered state, the linear stability analysis predicts that the homogeneous flocking state is destabilized by  splay fluctuations of the order parameter. In this case the instability is associated with spatial gradients in the direction normal to that the mean order. The wavevector of the fastest growing mode also scales  as $k_c\sim (v_0-v_c)^{1/2}$ at fixed density $\rho_0$ and as $(\rho_0-\rho_c)^{1/2}$ at fixed self propulsion speed, but it is clear that nonlinear pattern selection mechanisms must be involved  to yield the formation of the observed transverse stripes (associated with spatial gradients in the longitudinal direction) in this region.  On the other hand, the instability line obtained from the linear theory agrees remarkably well with the numerical onset of stripes in this high density region (Fig.~(\ref{fig:phasedia})). More work is needed to understand stripe formation at high density and the origin of the associated length scale.

\section{Nonlinear Regime}

To go beyond the linear stability analysis and investigate the nature of the flocking state above $v_c$, we have solved numerically the full nonlinear hydrodynamic equations. The numerical analysis has been carried out using the specific parameter values obtained for the self-propelled hard rod model of Ref.~\cite{aparna}, summarized in Table ~\ref{table1}.
\begin{table}
\centering
\begin{ruledtabular}
\begin{tabular}{c|c|c|c|c|c}
$D/D_0$ & $D_b/D_0$& $D_s/D_0$ &  $\lambda_1$ & $\lambda_2$ &$\lambda_3$  \\
\hline
$ (3+2v_0^2)/2$ &
$( 7+6v_{0}^{2})/8$ &
$(9+10v_{0}^{2})/8$ &
${3\pi v_{0}^{2}} $ & ${\pi v_{0}^{2}}$ &${\pi v_{0}^{2}} $
\end{tabular}
\end{ruledtabular}
\caption{Diffusion constants and convective parameters for the model of self-propelled hard rods with excluded volume interactions discussed in Ref.~\cite{aparna}. All diffusion constants are in units of $\ell^2D_r$ and all convective parameters are in units of $\ell^3D_r$. The diffusion coefficients have been expressed in terms of the longitudinal diffusion constant $D_0$ of a long, thin rod. Below we use the low density value, $D_0=1/4$.}
\label{table1}
\end{table}
All  diffusion coefficients  are enhanced by  self propulsion of an additive contribution proportional to $v_0^2$ that arises from the persistent nature of the random walk performed by Brownian, self-propelled rods.  In the following we discuss the properties of the system in terms of   two dimensionless parameters,
the self-propulsion speed, $v_{0}$, and the density of particles, $\rho_0 $.
In the numerics the coefficients $a_2$ and $a_4$ that control the continuous mean field phase
transition from an isotropic to a polar state have been  taken to be of the simple form given in Eqs.~\eqref{a2a4}
with $\rho _{c}=0.5$ in units of the rod length. This form yields $P_0\sim(\rho-\rho_c)^{1/2}$ for $\rho\rightarrow\rho_c$ and $P_0\rightarrow 1$ for $\rho>>\rho_c$.

For generality we include fluctuations beyond the mean field level in the  numerical analysis by adding Gaussian white noise terms in both the
density and polarization equations of the forms $\nabla \cdot \mathbf{f}_{\rho
}(\mathbf{r},t)$ and $\mathbf{f}_{P}(\mathbf{r},t)$, respectively. The random forces are chosen to have zero mean and correlations
\begin{equation}
<f_{i\rho }(\mathbf{r},t)f_{j\rho }(\mathbf{r}^{\prime },t^{\prime
})>=\delta _{ij}\Delta _{\rho }\rho (\mathbf{r},t)\delta (\mathbf{r}-\mathbf{
r}^{\prime })\delta (t-t^{\prime }),  \label{eq22}
\end{equation}
\begin{equation}
<f_{iP}(\mathbf{r},t)f_{jP}(\mathbf{r}^{\prime },t^{\prime })>=\delta _{ij}
\frac{\Delta _{P}}{\rho (\mathbf{r},t)}\delta (\mathbf{r}-\mathbf{r}^{\prime
})\delta (t-t^{\prime }),  \label{eq23}
\end{equation}
where $\Delta _{\rho }$ and $\Delta _{P}$ are dimensionless noise strengths.
The noise
in the density equation scales as $[\rho (\mathbf{r},t)]^{1/2}$, while the polarization noise scales as $[\rho (\mathbf{r},t)]^{-1/2}$
~\cite{sdean,shradhasr}. This difference arises because the fields $\rho $
and $\mathbf{P}$ are extensive and intensive quantities, respectively. The numerical results described below are all for fixed values of the noise amplitudes,   $\Delta _{p}=\Delta _{\rho }=0.3$.
We have solved the nonlinear equations using the Euler method for numerical
differentiation on a grid 
with $\Delta x=1.0$ 
and $\Delta t=0.1$ (we have verified that the numerical scheme is convergent and stable for $\Delta t / (\Delta x)^2<0.5$) 
We consider a square system of size $L\times L$ with both periodic and shifted boundary conditions and a range of system sizes.

The behavior of the system as a function of  the
self-propulsion velocity $v_{0}$ and the density of particles $\rho $ is
summarized in the phase diagram shown in Fig~\ref{fig:phasedia} discussed in the Introduction.
 The isotropic state is stable for all $v_0$ and  $\rho <\rho _{c}$. For $\rho>\rho_c$ and $v_0<v_c(\rho_0)$ the system is in the fluctuating flocking state, characterized by finite polarization and large spatial and temporal fluctuations of both density and order parameter.  For $v>v_c(\rho_0)$ we find a striped phase, with alternating ordered high density bands and disordered low density bands, propagating in the direction of order. In the numerics the value of $v_c$ is identified as the self propulsion velocity where the density histograms shown in Fig. ~(\ref{fig:histogram}) change from unimodal to bimodal. The histograms are constructed by recording the local density at each spatial grid point for fixed mean density $\rho_0$  averaged over many initial conditions. We have also verified that histogram of local polarization magnitude change from unimodal to bimodal as the same value of $v_0$. The numerical boundary for the onset of the stripe regime vanishes with $v_0$ for $\rho_0\rightarrow\rho_c$ and is in qualitative agreement with the boundary calculated in section    for the onset of the longitudinal instability. The theoretical curve, however, grows much faster with $v_0$ than the numerical boundary. Surprisingly, at high density the theoretical boundary for the linear instability of splay fluctuations agrees very well with the numerical onset of stripes.

\subsection{Fluctuating flocking state}

In this subsection we characterize the
fluctuating flocking state that exists in the region $\rho >\rho _{c}$ and $
v_{0}<v_{c}$ of the phase diagram in Fig.~\ref{fig:phasedia}. As noted earlier,
this state is characterized by large fluctuations in the density.
These fluctuations do not, however,
destroy the underlying orientational order of the system. This is displayed in Figs.~\ref{fig:OPtime} and \ref{fig:OPL}.
\begin{figure}[tbp]
\begin{center}
{\includegraphics[height=9.0cm, width=9.0cm,angle=270]{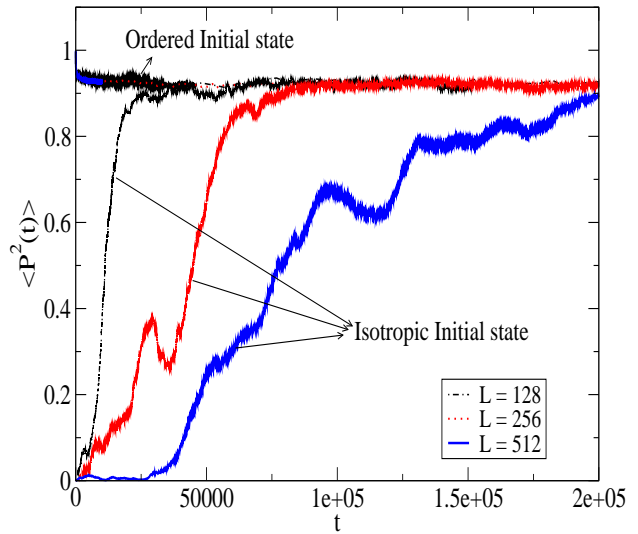}}
\end{center}
\caption{(color online) The  magnitude squared of the orientational order parameter $
\langle P ^2(t)\rangle$ as
a function of time for an isotropic initial state ($\langle P^2(t=0)\rangle=0$) and an ordered initial state ($\langle P^2(t=0)\rangle=1$), for three system sizes and $v_0=0.1$, $\rho_0=0.7$, corresponding to $v_c\sim 0.42$. The three curves obtained for an isotropic initial state overlap and cannot be distinguished in the figure. Both initial states approach the same
macroscopically ordered state at long time, although the time scale required for the  isotropic initial state
to reach the asymptotic steady value is much longer.}
\label{fig:OPtime}

\end{figure}
Figure~\ref{fig:OPtime} shows the time evolution of the magnitude squared of the order parameter, $\langle P^2(t)\rangle=\langle P_x^2(t)+P_y^2(t)\rangle$, where here and below the brackets denote a spatial average over the system and an average over different realizations of initial conditions,  for both an initial ordered ($\langle P^2(t=0)\rangle=1$) and an initial disordered ($\langle P^2(t=0)\rangle=0$) state. Both states reach the same ordered state at long times, although on very different time scales.
\begin{figure}[tbp]
\begin{center}
{\includegraphics[height=9.0cm, width=9.0cm]{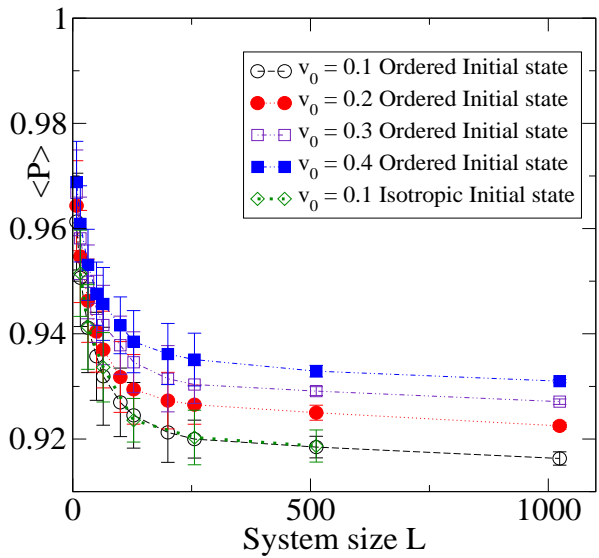}}
\end{center}
\caption{(color online) The mean magnitude of the polar order parameter as a function of system size $L$
for
different values of $v_0$ and $\rho_0=0.7$, corresponding to $v_c \simeq 0.42$. The curves appear to saturate
as the system size increases, suggesting that the system
remains macroscopically ordered in the thermodynamic limit.}
\label{fig:OPL}
\end{figure}
The asymptotic state is ordered (i.e., $\langle P^2 \rangle \neq 0$)and this does  not appear to be an	 artifact of the finite system size, as shown in Fig.~\ref{fig:OPL}
where the magnitude of the order parameter is displayed as function of system size for various values of $v_0$ and different initial states. The numerics suggest that $P$ is finite and close to unity (note the narrow range of $\langle P\rangle$ on the vertical axis of Fig.~\ref{fig:OPL}) in the fluctuating flocking state.
The large difference in the relaxation time from an initial disordered/ordered state to the asymptotic ordered state seen in Fig.~\ref{fig:OPtime} is not unexpected. When starting in a disordered state, the system
locally finds different degenerate ordered states, which
subsequently coarsen towards the homogeneous ordered state. Below we characterize both the coarsening
behavior as the system seeks out its asymptotic steady state and
the properties of the asymptotic fluctuating flocking state.

To quantify the coarsening behavior we have measured
the two-point correlation function of both the density and the order
parameter, defined as
\begin{eqnarray}
&&C_{\rho }\left( \mathbf{r},t\right) =<\delta \rho (\mathbf{r_{0}}
+\mathbf{r},t)\delta \rho (\mathbf{r_{0}},t)>\;,\label{Crho}\\
&&C_{P}\left( \mathbf{r}%
,t\right) =<\mathbf{P}(\mathbf{r_{0}}+\mathbf{r},t)\cdot \mathbf{P}(
\mathbf{r_{0}},t)>\;.
\end{eqnarray}
Before discussing the behavior of these correlation functions, it is useful
to recall the  dynamics of phase ordering  developed in the context of
equilibrium second order phase transitions \cite{phaseordering}. Phase ordering theories consider a  system in an initially disordered state that is rapidly quenched below the order-disorder transition point and describe the time evolution following the quench. Immediately after the quench the system consists of finite-size ordered regions, each in one of the continuum of degenerate ground states that correspond to one choice of the spontaneously broken continuous symmetry. The system then evolves in time and  ``coarsens" with some of the ordered regions growing at the expense of others and eventually taking over the entire system. The coarsening process  is typically controlled by a single energy scale, namely the energy cost
of the domain walls between different ordered regions. This implies that the time evolution of the system occurs via the growth of a single length scale $L(t)$ that
characterizes the size of a typical ordered region in the system
at a time $t$. The
 order parameter correlation function will then depend on time only through $L(t)$,
i.e., $C_{P}\left( r,t\right) =C\left( \frac{r}{%
L\left( t\right) }\right) $. Scaling analysis indicates that the length scale $L(t)$  grows with the dynamical critical exponent $z$,
$L\left( t\right) \sim t^{1/z}$. For a vector order parameter in two dimensions in equilibrium
one expects  $z\sim 2$~ \cite{phaseordering}.

In the case of  self
propelled particles, the orientational fluctuations that drive the coarsening of the ordered state also induce mass fluxes
and hence  couple to density fluctuations.
As a result, density correlations are essentially slaved to
the order parameter correlations and both $C_{\rho }$ and $C_P$ are expected to
exhibit coarsening behavior \cite{dasbarma}. This is indeed the behavior that has been observed in  active nematic liquid crystals, where both density and orientational correlations have been shown to coarsen on a characteristic length scale that grows like $t^{1/z}$, with $z\sim2$~\cite
{shradhaGinellietal}.
\begin{figure*}[tbp]
\begin{center}
\subfigure{\includegraphics[height=8.0cm, width=8.0cm,angle=270]{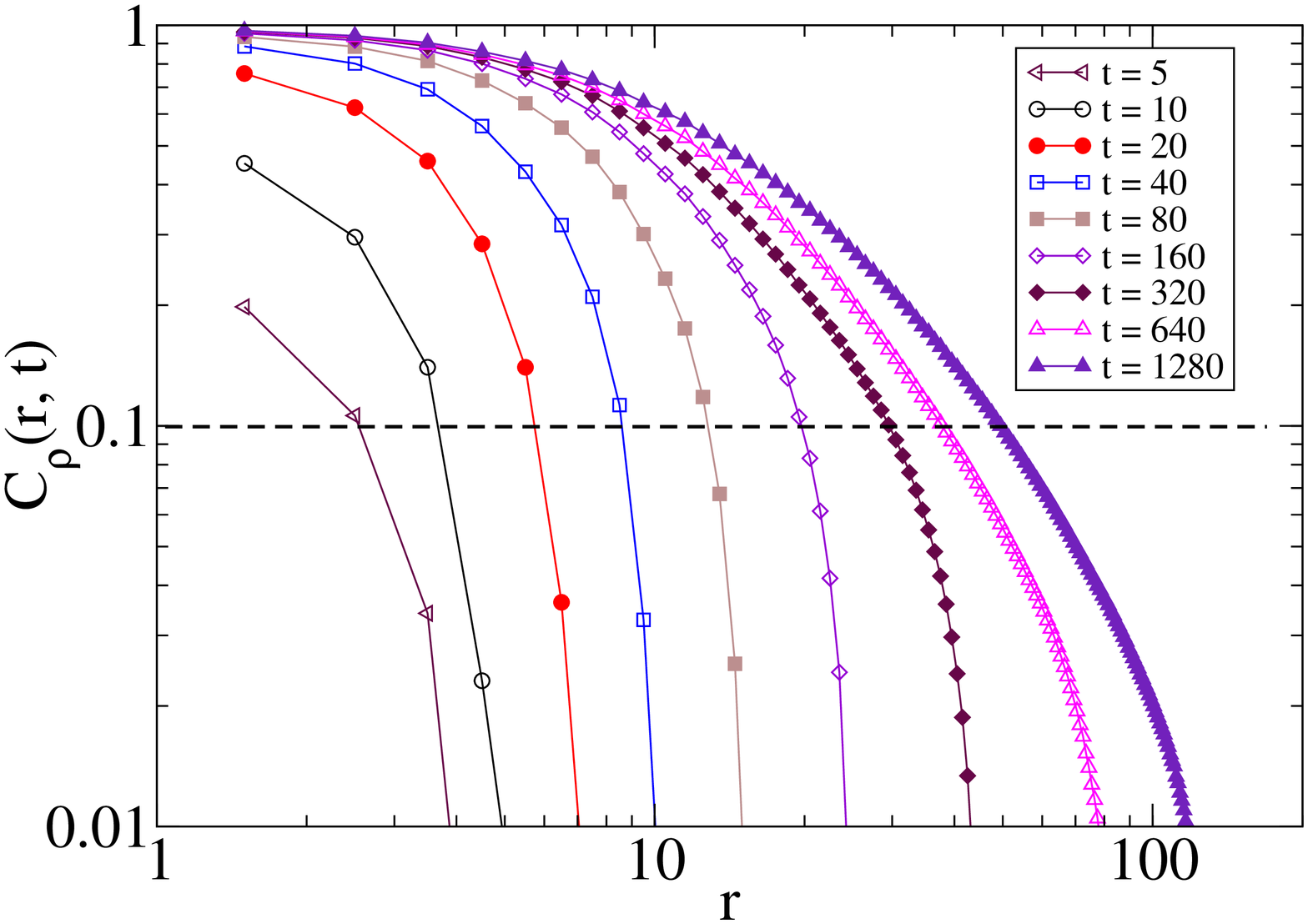}}
\subfigure{\includegraphics[height=8.0cm, width=8.0cm,angle=270]{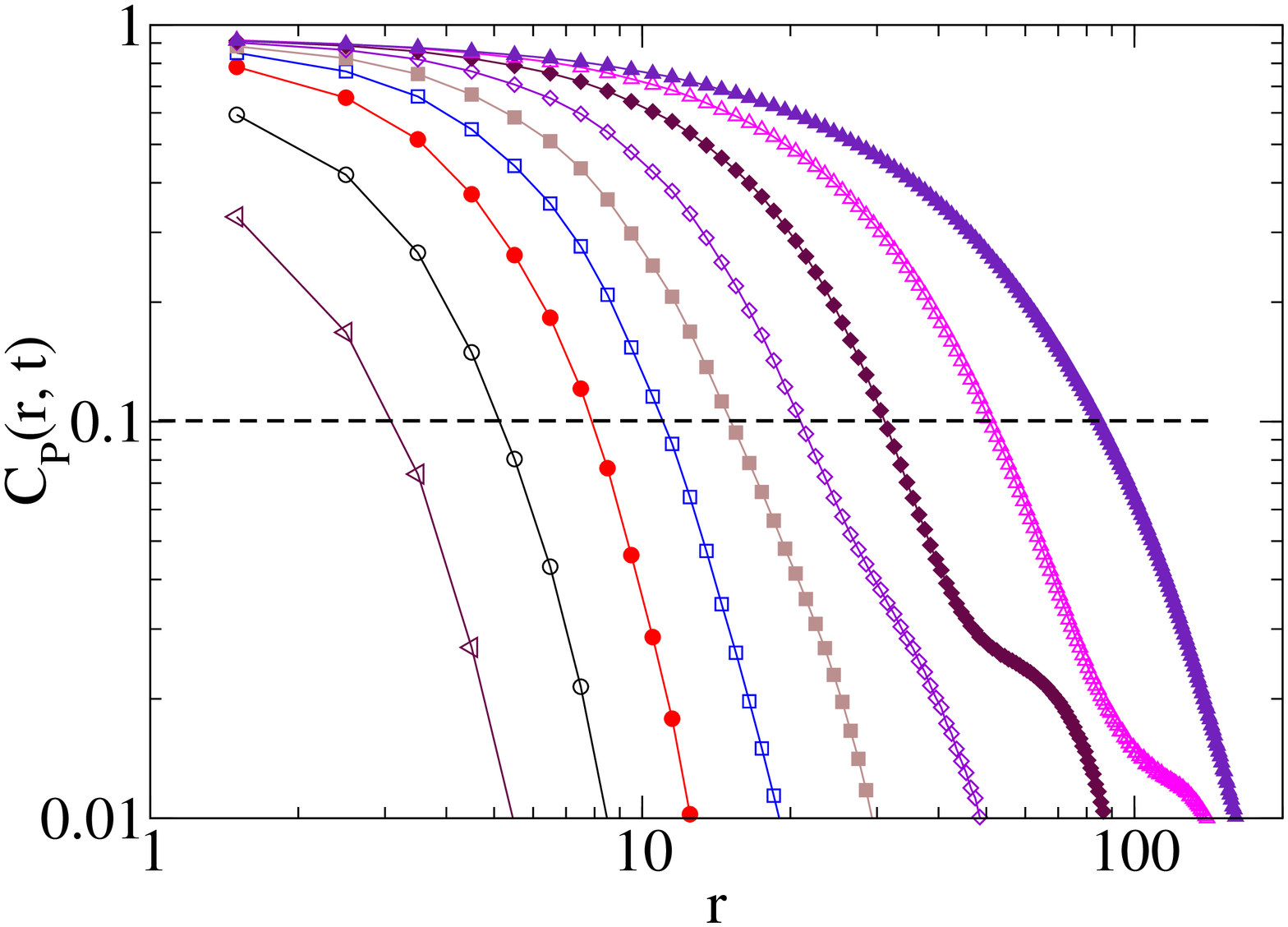}}
\end{center}
\caption{(color online) Early time,  two-point density
(top panel) and order parameter (bottom panel) correlation
function for $v_{0}=0.1 < v_c = 0.42$ mean density $\protect\rho _{0}=0.7$, and
$L=1024$. The dashed horizontal line indicates the value
of the correlation function at which we extract the coarsening
length scale $L(t)$, shown in Fig. \ref{fig:lengthiso}. }
\label{fig:corriso}
\end{figure*}

The behavior of {\it polar} active system appears to be somewhat different.
Figure~\ref{fig:corriso} shows the  two-point correlation function for the density and the
order parameter for our flock. Both exhibit a growing
correlation length as a function of time, but they do not exhibit the simple scaling
behavior outlined above, indicating that the approach to the homogeneous state is
no longer controlled  by the single energy scale associated with the cost of a domain wall.
Convective fluxes induced by self-propulsion
lead to correlations on longer length scales and hence accelerate the
coarsening dynamics. This picture can be substantiated by extracting a
length scale $L\left( t\right) $ from the correlation functions.
This is shown in Fig.~\ref{fig:lengthiso} where we see that the dynamical exponent $z$ is smaller than the equilibrium value, indicating that the coarsening dynamics in polar active systems  is faster than that of both equilibrium systems with vector order parameters and active
nematics. This may be due to the fact that, in a polar self-propelled
system, there is true long range order in 2D~\cite{tonertuPRL95,tonertuPRE98} and the
associated suppression of the Goldstone mode by the nonlinear couplings
changes the dynamics of the system as the orientational order builds up
towards the homogeneous ordered state.
\begin{figure}[tbp]
{\includegraphics[height=9.0cm, width=9.0cm,angle=270]{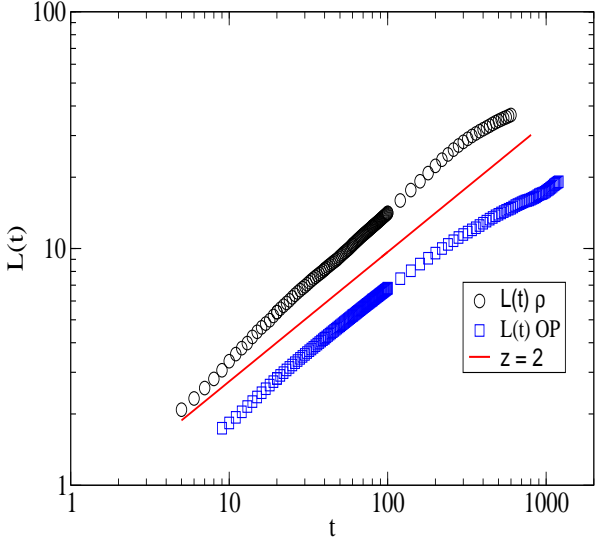}}
\caption{(color online) The coarsening length $L(t)$ as a function of time for the
density (circles) and order parameter (squares) correlations for $v_0=0.1$, $\rho_0=0.7$ and $L=1024$. For this density $v_c \simeq 0.42$ and the system was started in a disordered state.
The straight line has slope 2.
Although no single scaling exponents can be extracted for the two length scales $L(t)$, it is clear
that the growth with time is faster than would be obtained for  $z=2$. }
\label{fig:lengthiso}
\end{figure}

At long times, for $v_0<v_c$, the system reaches the ``fluctuating flocking state", a steady state
with finite mean polarization and anomalous fluctuations. To characterize the properties of this state we have evaluated  the two-point correlation function of
\emph{fluctuations} in the orientational order  parameter,
\begin{equation}
C_{\delta P}\left( \mathbf{r}
,t\right) = <\delta \mathbf{P}(\mathbf{r_{0}}+\mathbf{r},t)\cdot \delta
\mathbf{P}(\mathbf{r_{0}},t)>,
\label{cdeltap}
\end{equation}
 with $\delta{\bf P}({\bf r},t)={\bf P}({\bf r},t)-P(t)$. This is shown in  the right frame of Fig.~\ref{fig:corrst} and it decays
logarithmically as expected for vector order in 2D. The correlation function shown in Fig.~\ref{fig:corrst} has been averaged over ${\bf r}_0$, hence it represents only the isotropic (angular-averaged) part of the order parameter correlations. In general we expect the correlation function to be anisotropic and its spatial decay to be described by different length scales in the directions longitudinal and transverse to the direction of mean motion, as described in Ref.~\cite{tonertuPRE98}.  We have calculated the spatial decay of the order parameter correlation in each direction and find it indeed to be anisotropic, as shown  in Fig.~(\ref{fig:corr_aniso}). The theoretical analysis of Ref.  \cite{tonertuPRE98} predicts a power law behavior with different exponents characterizing the decay along and orthogonal to the direction of broken symmetry. Given the large spatial {\em and temporal} fluctuations in our system, the system sizes considered here are too small to obtain reliable statistics to quantify this behavior and extract scaling exponents.

\begin{figure}[tbp]
{\includegraphics[height=9.0cm, width=9.0cm,angle=-90]{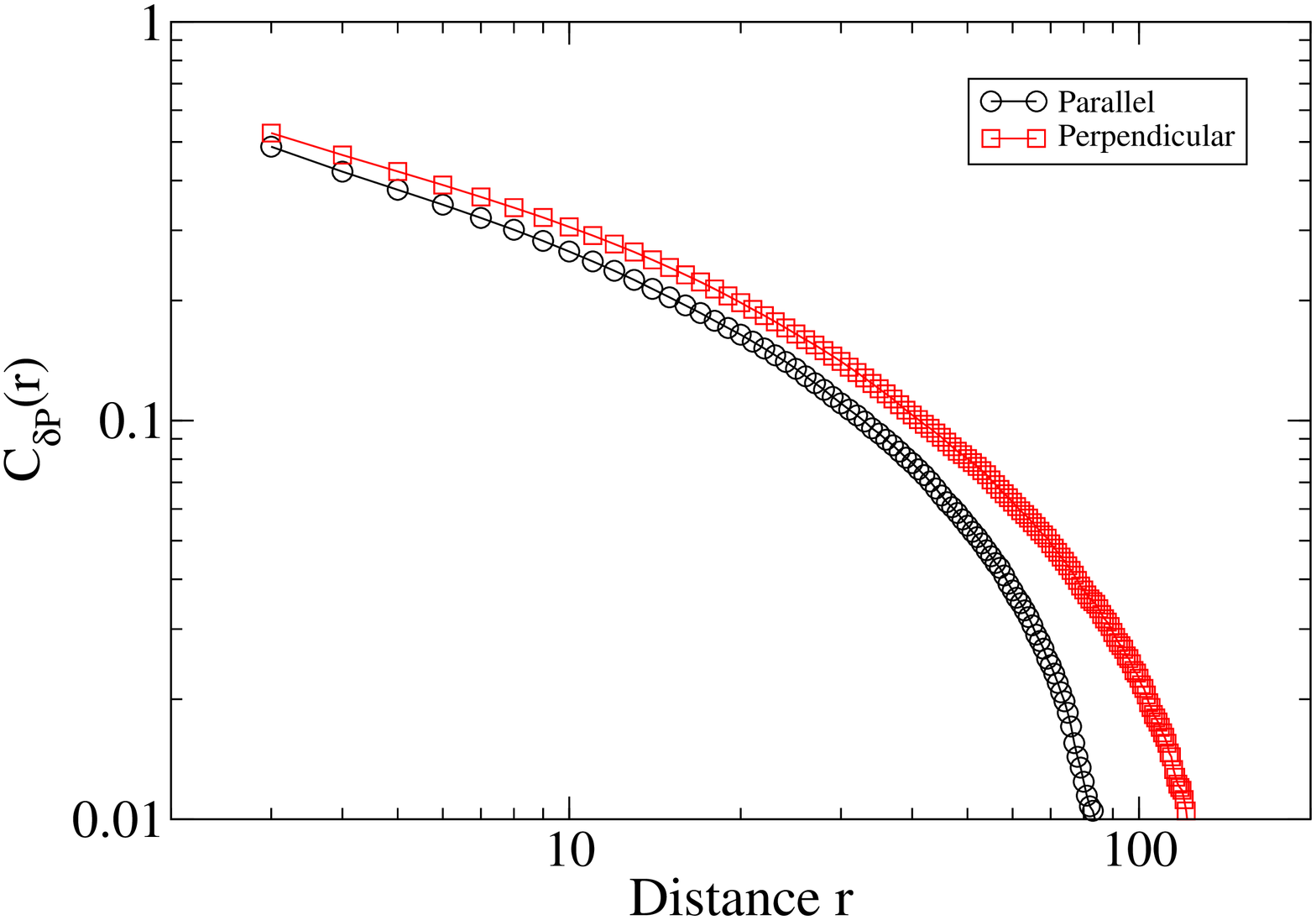}}
\caption{(color online) Plot of $C_{\delta P}\left( \mathbf{r}
,t\right)$ for the case when ${\bf r}$ is parallel to $\hat{p}_0$ (black/open circles) and when ${\bf r}$ is perpendicular to it (red/ filled circles). The anisotropy of the correlation functions begins to appear at large $r$, but larger system sizes are needed to quantify the difference.}
\label{fig:corr_aniso}
\end{figure}

In contrast to the order parameter correlations, which decay logarithmically,
the two point density correlation shown in the in the left frame of Fig.~\ref{fig:corrst} displays correlations over longer length scales than expected in equilibrium.  In fact the density correlation functions
exhibits cuspy  behavior of the form $C_{\rho }\left(
r,t\right) =1-\left( \frac{r}{L\left( t\right) }\right) ^{\alpha }$ with $%
\alpha \simeq 0.6$ typically characteristic of a state with growing domains. Furthermore, $C_{\rho }\left( \mathbf{r},t\right) $
depends very weakly on the self propulsion speed. These large correlations
in the density arise because of the coupling of this conserved field to the
fluctuations in the underlying order parameter field that is an intensive
variable. This leads to what has been termed ``giant number fluctuations''
in these active systems \cite{RamaswamyEPL} and is the underlying mechanism
for the large fluctuations in the density in the flocking state of our
system.

\begin{figure*}
\begin{center}
\subfigure{\includegraphics[height=8.0cm, width=8.0cm,angle=270]{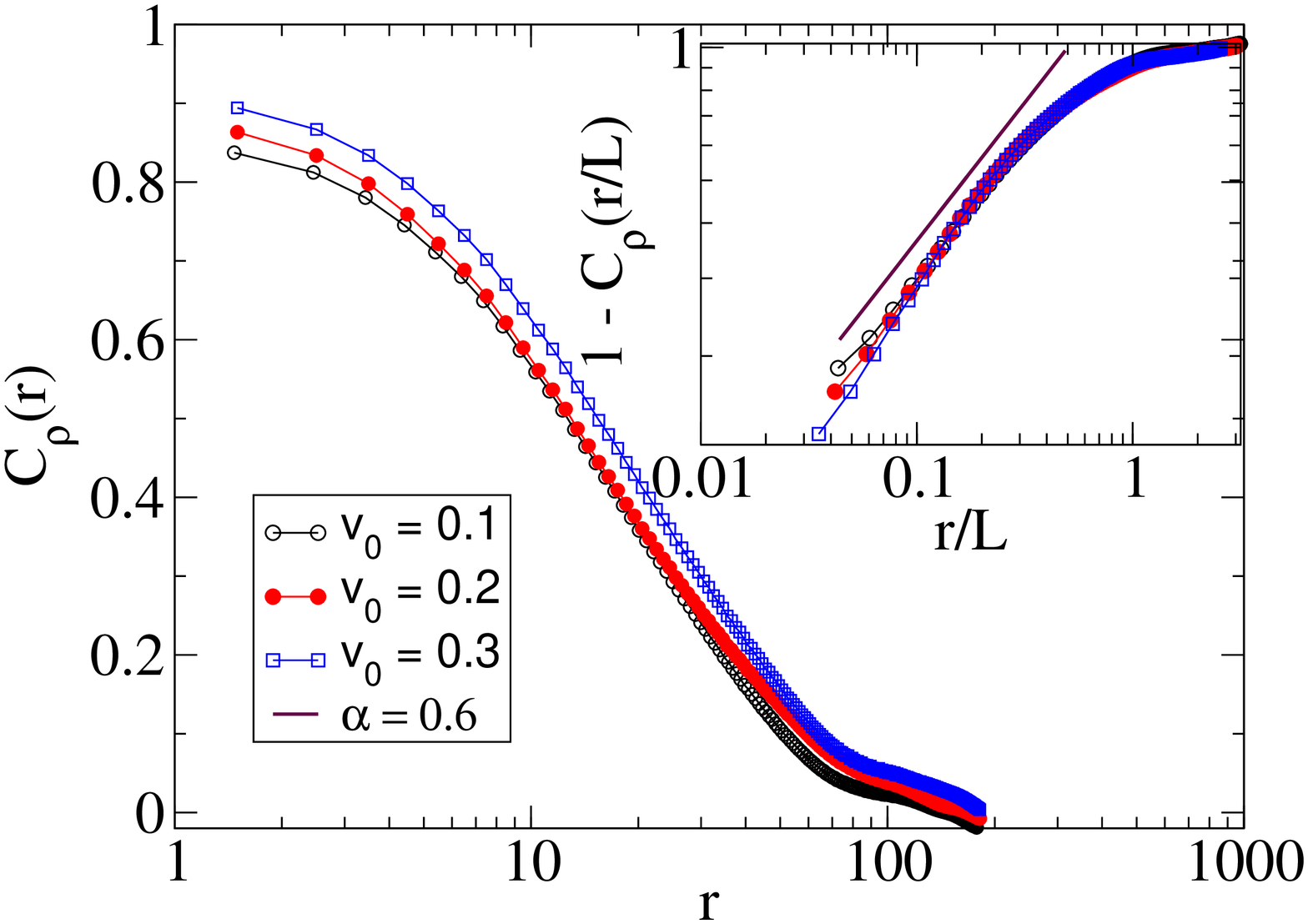}}
\subfigure{\includegraphics[height=8.0cm, width=8.0cm,angle=270]{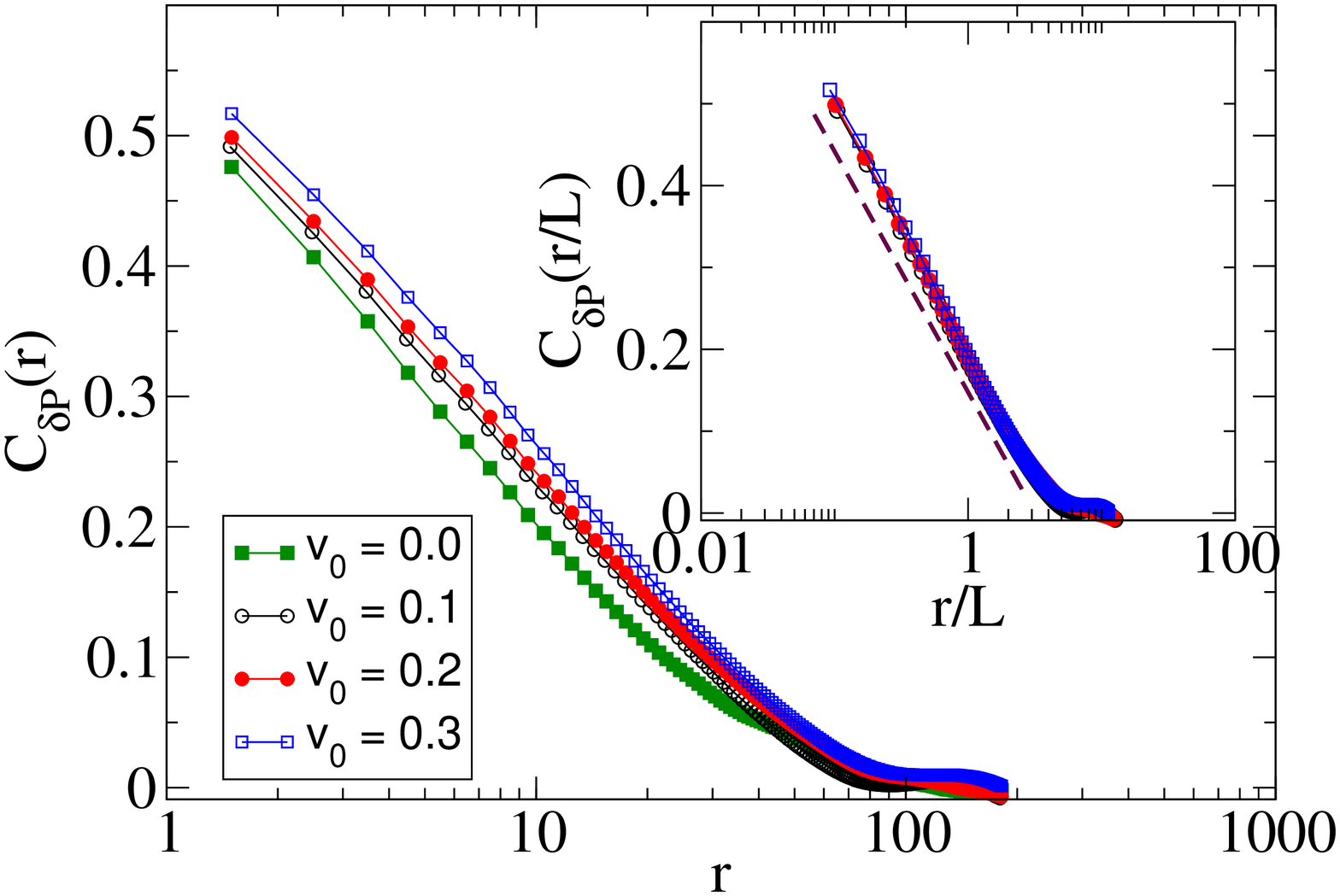}}
\end{center}
\caption{(color online) Late-time two point correlation functions of  density, $C_{\rho}(r)$, and  order
parameter, $C_{\delta P}(r)$, fluctuations. The data are obtained from an
initial ordered configuration for $\rho_0=0.7$, $L=512$ and various values of $v_0$, in the region corresponding to the fluctuating flocking  state. Inset (top panel) shows $(1-C_{\rho}(r/L(t)))$ vs. the scaled distance $r/L(t)$, for
various values of $v_0$.
The dashed straight line shows the cuspy nature of two-point correlation function
with cusp exponent $\alpha = 0.6$. Inset (bottom panel) shows the scaled
two-point  order parameter correlation function, $C_{\delta P}(r/L(t))$, vs. the scaled distance $r/L(t)$, for several $v_0$. The dashed straight line shows the logarithmic
decay of $C_{\delta P}(r/L(t))$.}
\label{fig:corrst}
\end{figure*}

\subsection{ Striped Phase}

 The top inset of Fig.~\ref{fig:phasedia}
shows a real space snapshot of the density obtained for $\rho_0>\rho_c$ and $v_0>v_c(\rho_0)$ In this region the systems consists of well defined stripes
of the high density ordered phase alternating with stripes of the low density disordered phase.
In the ordered region the polarization is always normal to the long direction of the stripes and the stripes travel at a fixed speed in the direction of polar order.  Panels (a) and (b) of Fig.~\ref{fig:histogram} show histograms of the density and magnitude of the order parameter
for a fixed  density $\rho_0=0.7>\rho_c$ and different
 self propulsion speeds, $v_{0}$.
 For small $v_{0}$, the
histograms are unimodal
(fitted with a gaussian peaked at mean density $\rho_0 = 0.7$), signalling  a uniform state. Above a characteristic value of  $v_{0}$ the histograms acquire a bimodal structure (fitted with two overlapping Gaussians peaked at low and high densities), corresponding to the striped phase. The boundary $v_c(\rho_0)$ corresponding to the onset of the striped phase and shown in Fig.~\ref{fig:phasedia} is determined as the value of $v_0$ corresponding to the onset of this bimodal structure. The error bars on these data point are simply  the step size of our increments in $v_0$. Within these error bars,  the same values of $v_c(\rho_0)$ are obtained from  the onset of a bimodal structure in both the density and polarization histograms. The points shown in the phase diagram are  obtained using the density histograms.
 As noted earlier, this  boundary closely tracks
the threshold line for the onset of the linear instability of splay fluctuations at large $\rho_0 $ (although this instability arises from spatial gradients normal to those of stripe formation!),  and vanishes with $v_0$ as $\rho_0\rightarrow \rho_c^+$, as predicted by the longitudinal linear instability discussed in Section III. However, there is considerable discrepancy between the linear theory and the numerics in the details of the behavior at low density. 
%
\begin{figure*}[tbp]
\begin{center}
{\includegraphics[height=15.0cm, width=15.0cm, angle=270]{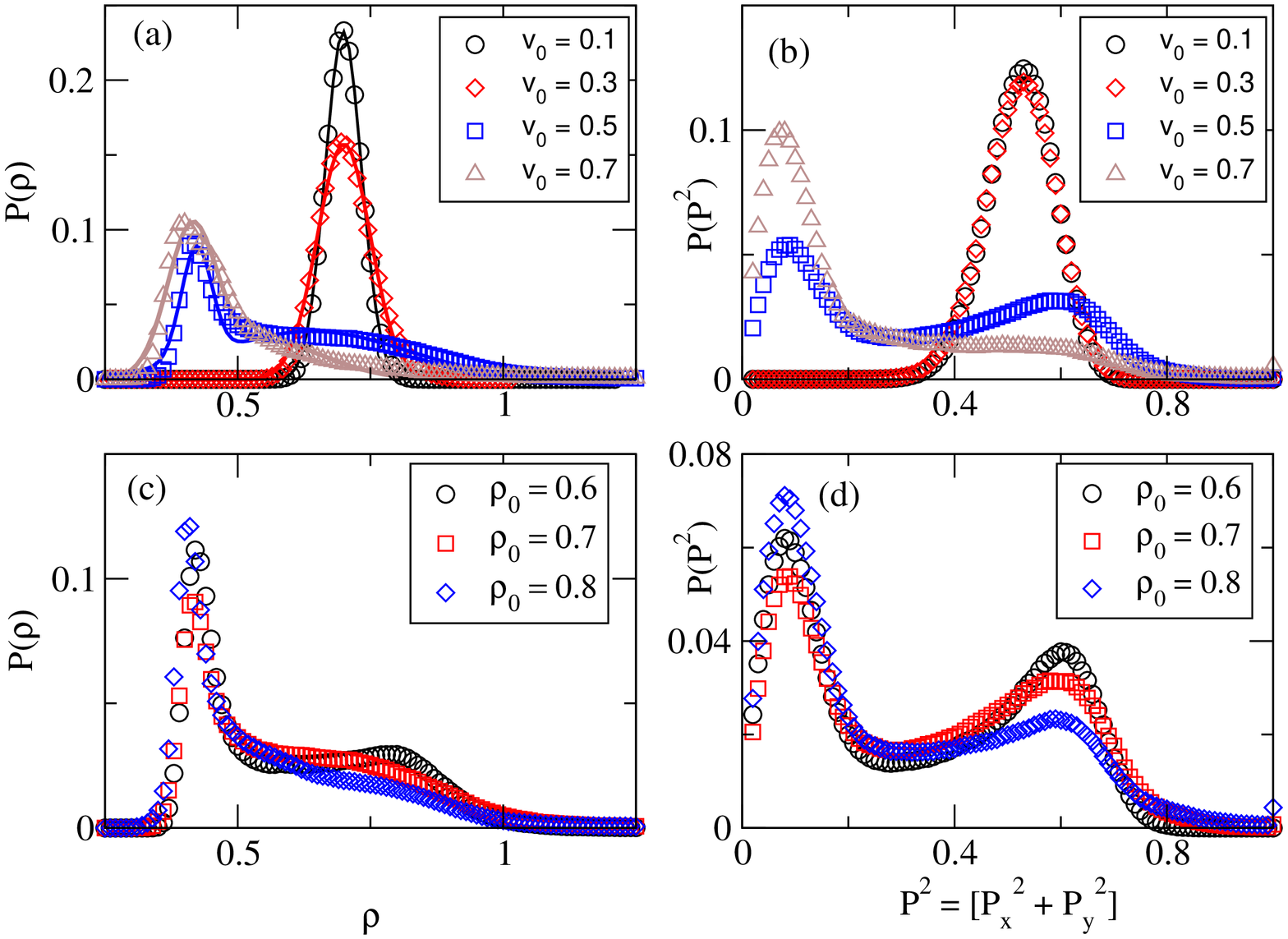}}
\end{center}
\caption{(color online) (a)-(c) Density and (b)-(d) order parameter histograms. In panels
(a) and (b) the histograms are shown for  $\protect\rho_0 =
0.7 $,  $v_0 = 0.1$, $0.3$, $0.5$, $0.7$ and  $L = 128$. For
small $v_0$, below the threshold value $v_c \simeq 0.42$, the histograms are unimodal (fitted by a Gaussian),
indicating a uniform state. For $v_0>v_c$, the histograms are bimodal (fitted by two Gaussian curves), indicating
the onset of stripes. In panels (c) and (d) the histograms are shown for $%
v_0 > v_c$ and three values of the mean density, $\protect\rho_0 = 0.6$, $0.7$, $%
0.8$. The position of the peaks does not change with density, while the difference in
the height of the two peaks increases with increasing density.}
\label{fig:histogram}
\end{figure*}

The bottom two panels of Fig.~\ref{fig:histogram} show histograms of density and polarization for a fixed
self propulsion speed $v_{0}>v_{c}$ and three different values of density. These histograms are used to infer the properties of this striped phase. The difference in position of the two peaks in the bimodal histograms indicates the contrast in density and order parameter between the ordered and disordered stripes. The position of the peaks is independent of the self propulsion speed and only weakly dependent on the density, suggesting that the density contrast between the isotropic and ordered stripes is entirely diffusion limited. The height of the high density/high order peaks in the bimodal histograms is a measure of the width of the ordered stripe with respect to the disordered one. We note from the figure that the height decreases with both increasing $v_0$ and $\rho_0$. This indicates that the width of the ordered stripe decreases with increasing values of these two parameters.
Further, we measure the speed of propagation of the stripes, shown in Fig.~\ref{fig:velocitywidth}. As naively expected, the propagation speed of the stripes increases linearly with $v_0$.
\begin{figure}[tbp]
\begin{center}
{\includegraphics[height=9.0cm, width=9.0cm]{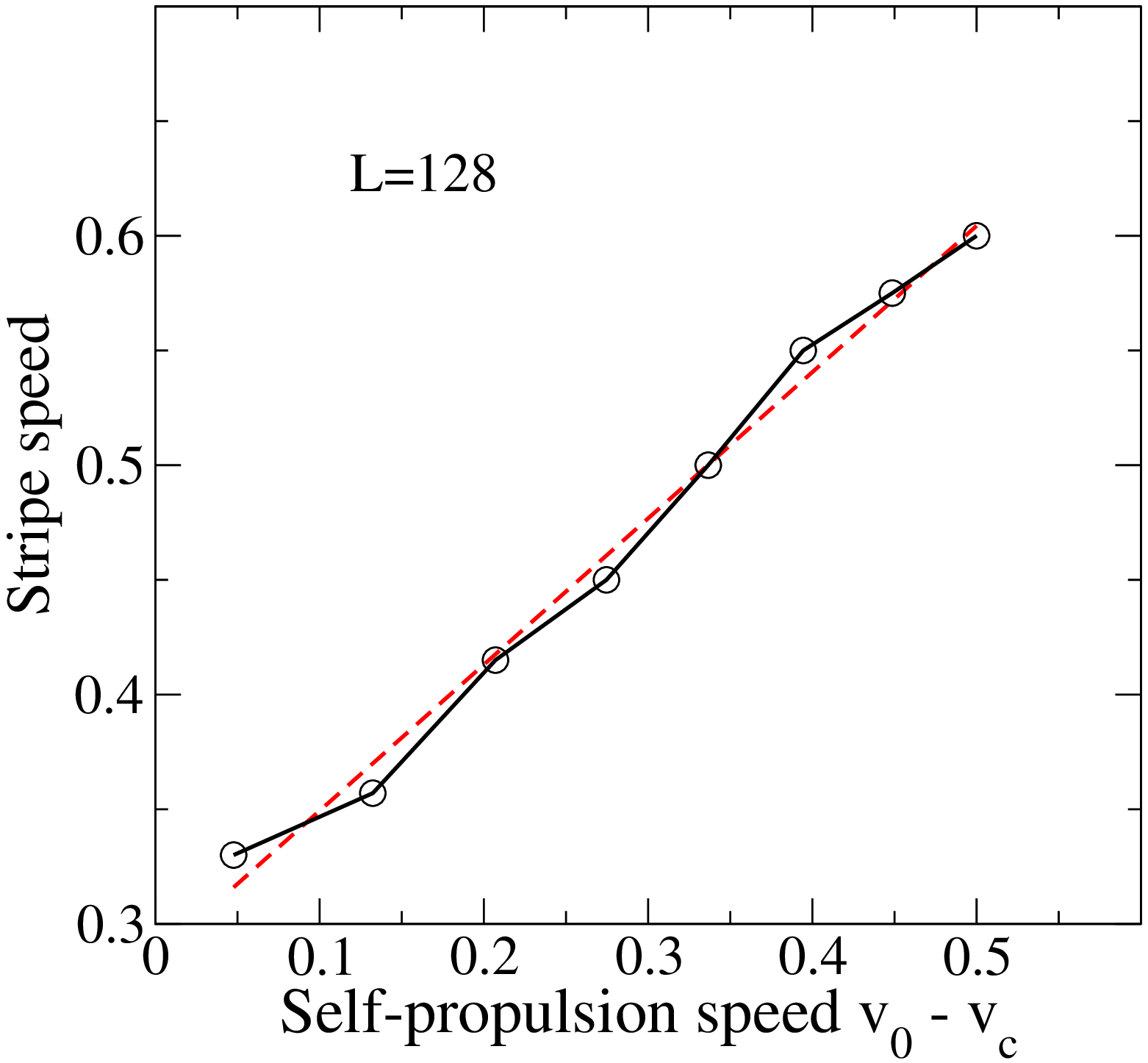}}
\end{center}
\caption{(color online) Speed of the stripes as a
function of $v_0 - v_c$ for $\rho_0=0.7$.
The speed of propagation of the stripes increases linearly
with $v_0$. The solid line is a guide to the eye. The dashed line is a linear fit  $\sim v_0-v_c$}
\label{fig:velocitywidth}
\end{figure}

An alternative way of displaying the existence of the stripes and quantifying their properties is
provided by the two point density
correlation function defined in Eq.~(\ref{Crho}). We have evolved the system starting form a uniform  ordered  initial state and evaluated the two-point correlation function  as a function of ${\bf r}$ for three directions:
$0^{\circ },$ $%
45^{\circ }$ and $90^{\circ }$ to the direction of initial orientational order. The result is shown in Fig.~\ref{fig:directionalstripescorr}.
 When
the self propulsion speed $v_{0}<v_{c}$ (right frame), the correlation decays
monotonically in all directions. On the other hand, when $v_{0}>v_{c}$ (left frame), we
find well defined oscillations in the correlation function in the directions normal and at $45^{\circ }$ to the direction of motion showing the
emergence of the periodic structure associated with the stripe pattern.
\begin{figure}[tbp]
\begin{center}
\subfigure{\includegraphics[height=8.0cm, width=8.0cm, angle=270]{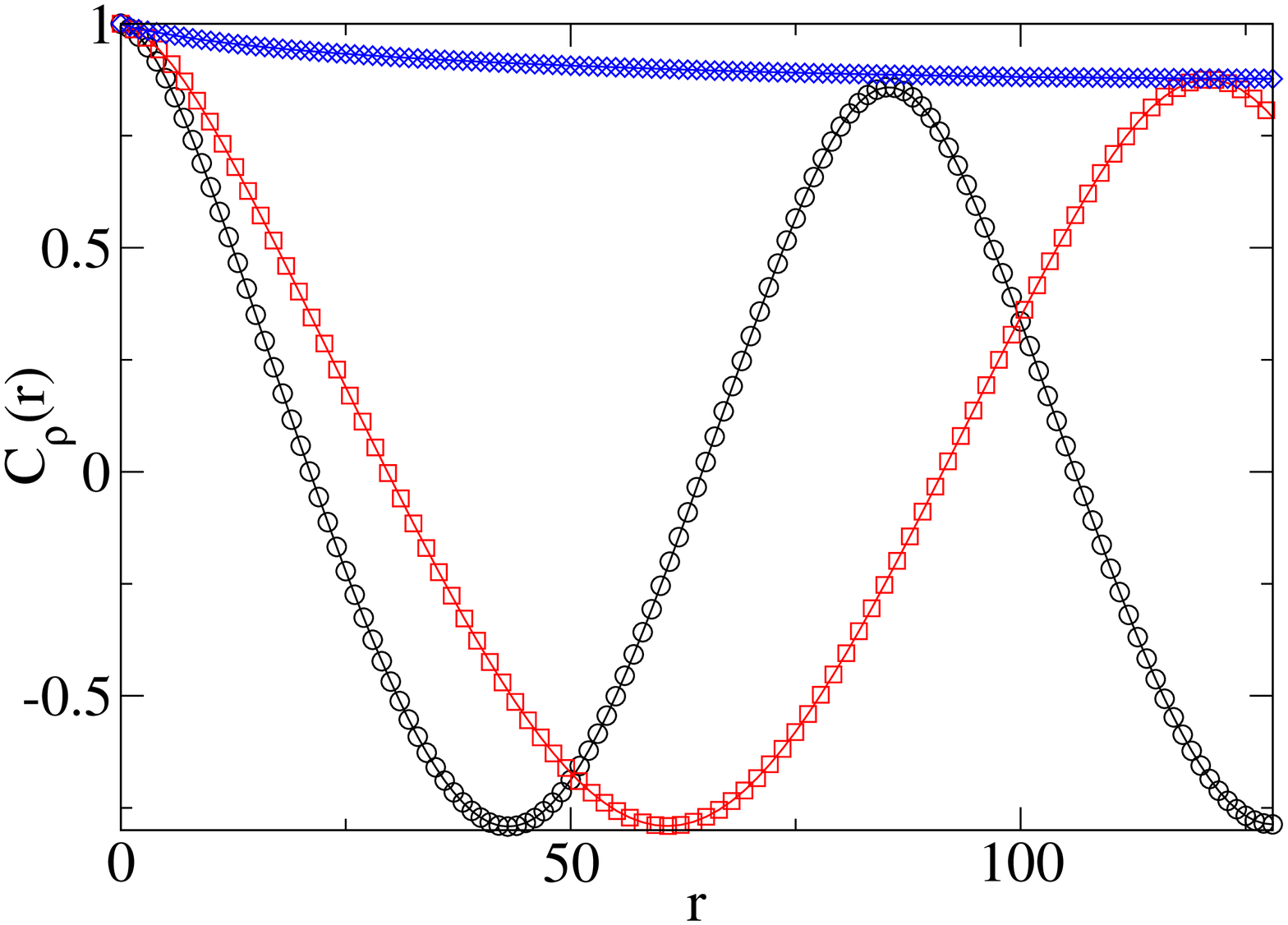}}
\subfigure{\includegraphics[height=8.0cm, width=8.0cm, angle=270]{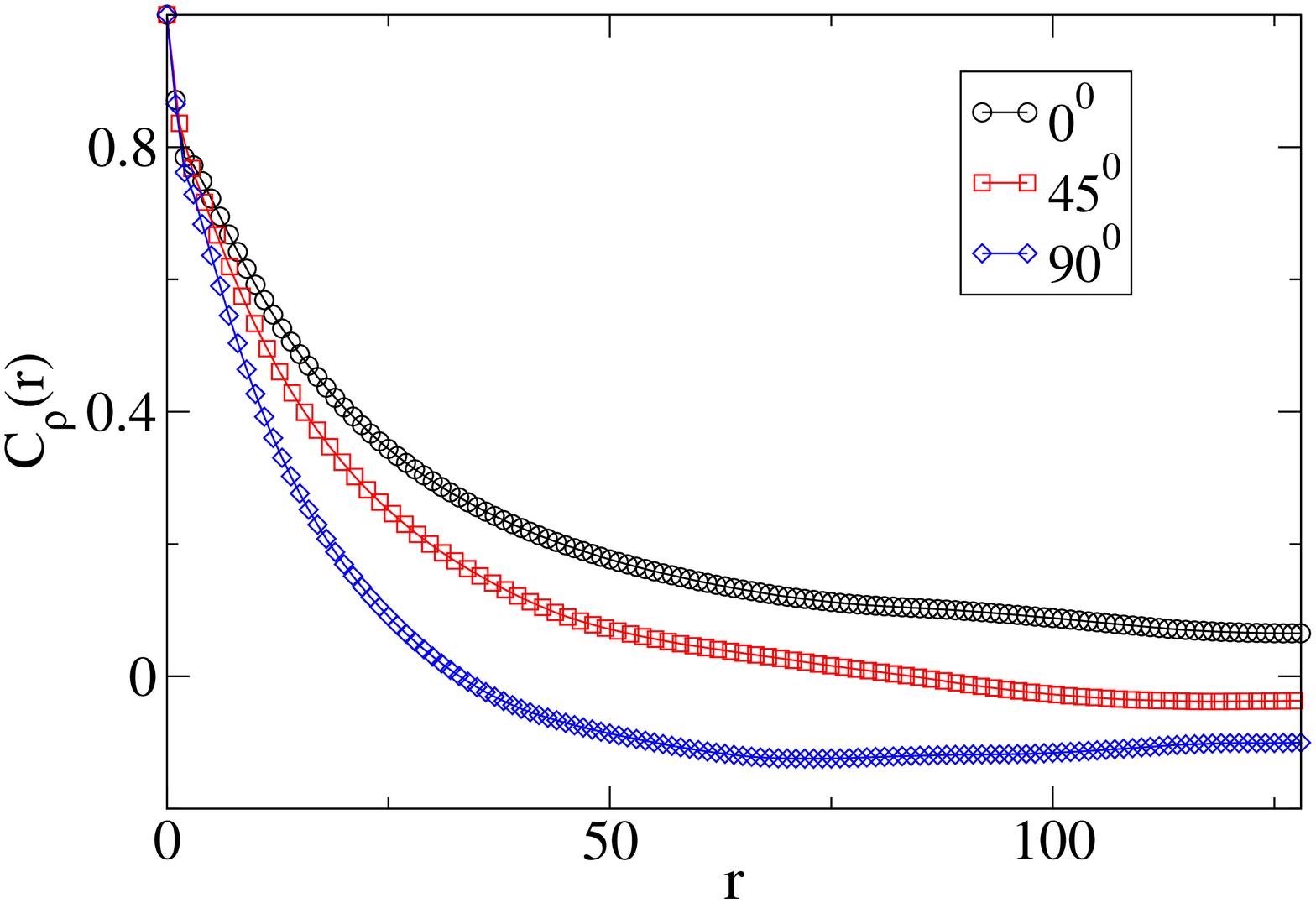}}
\end{center}
\caption{(color online) Steady-state two-point density correlation function for three
different directions with respect to the ordering direction, $90^{\circ}$, $
45^{\circ}$ and $0^{\circ}$, for $v_0 > v_c$ (top panel) and for $v_0 < v_c$ (bottom panel).
For $v_0 < v_c $, there is no directional dependance in the correlation
function. For $v_0 > v_c$, correlations at $90^{\circ}$ to the ordering direction decay monotonically,
while correlations at
$0^{\circ}$ and $45^{\circ}$ to the ordering direction show
oscillations. }
\label{fig:directionalstripescorr}
\end{figure}
Also, this clearly indicates that the spatial inhomogeneity develops in the
direction of initial orientational ordering  even in the region where the linear
instability is along a wavevector orthogonal to the ordering direction. 
Finally we have also investigated the dependence of stripe formation  on system size and boundary conditions. We find that the width and speed of the
stripes remain mostly invariant as we go to larger system sizes. We have also
solved our equations with shifted boundary conditions \cite{shiftedaldana}
and have found that the striped phase persists.
%
%

In summary, for values of $v_0>v_c(\rho)$ given by the (black) solid line in Fig.~(\ref{fig:phasedia}),  the system develops robust propagating stripes of alternating ordered and disordered regions. The numerically identified transition line follows closely the threshold for the splay instability identified in Eq.~(\ref{eqvc}), goes to zero at the phase transition in agreement with the longitudinal instability identified using Eq.~(\ref{deff}) and shows a behavior unlike both of these linear instabilities in the intermediate region. Further, for systems initialized in a uniform ordered state, these stripes form along the direction of initial ordering even in the domain where the longitudinal instability is absent. This suggests that the pattern selection  arises from a complex interplay between the unstable linear modes \cite{crossRevmodphys}. Also, the width of the stripes exhibits a scaling behavior consistent with the critical wavevectors $k_c$ of both instabilities, but is not quantitatively captured by either length scale. A systematic study of the relationship between the linear modes and the patterns observed here  will be the focus of a future work.

We can show that the  nonlinear equations admit a propagating front solution that may correspond to the onset of the stripe phase, although the stability if this solution is yet to be established. In our numerical study we have solved the equations (\ref{nonlinDens}) and (\ref{nonlinPol}) by systematically dropping various nonlinear terms. We have established that the terms that are critical for the formation of the striped phase are the homogeneous nonlinearity in the coefficient $a_2$ that induces the phase transition, the couplings between density and polarization embodied by the convective terms in the density and polarization equations ($v_0\bm\nabla\rho{\bf P}$ and $v_{0}(\bm\nabla
\rho)/\rho $, respectively)
and the convective nonlinearity controlled by the parameter $\lambda _{3}$.
The longitudinal instability arises from the interplay of $a_2$ and the convective terms while the splay instability is controlled by the term proportional to $\lambda_3$. It is useful then to consider a simplified description of the nonlinear dynamics where diffusion is neglected and only terms essential for  the pattern formation
are retained, given by
\begin{equation}
\partial _{t}\rho =-\bm\nabla \cdot (\rho v_{0}\mathbf{P})
\label{rho-simple}
\end{equation}
\begin{eqnarray}
\partial _{t}\mathbf{P}
&=&-\left[ a_{2}\left( \rho \right) +P^{2}a_{4}\left( \rho \right) %
\right] \mathbf{P}  \notag \\
&&-\frac{v_{0}}{2\rho }\bm\nabla \rho +\lambda _{3}P_{i}\bm\nabla \rho P_{i}
\label{P-simple}
\end{eqnarray}
%
Denoting by $x$ the direction of broken symmetry of the putative HP state, we postulate a solution of these equations in the form of a front uniform in $y$ and propagating along $x$ with a
yet undetermined constant speed $U$,
\begin{eqnarray}
&&\rho \left(
x,y,t\right) =\rho \left( x-Ut\right), \\
&&\mathbf{P}\left( x,y,t\right)
=P\left( x-Ut\right) \mathbf{\hat{x}}\;.
\end{eqnarray}
Inserting this ansatz, Eqs.~(\ref{rho-simple}) and (\ref{P-simple}) become
\begin{equation}
\partial _{x}\ln \rho =\frac{v_{0}}{\left( U-v_{0}P\right) }\partial _{x}P\;,
\label{densSimple}
\end{equation}
\begin{equation}
\left( a_{2}+a_{4}P^{2}\right) P-\left( \lambda _{3}P^{2}-\frac{v_{0}}{
2\rho }\right) \partial _{x}\rho -\left( U+\lambda _{3}\rho P\right)
\partial _{x}P=0.  \label{polSimple}
\end{equation}
The density equation can be formally integrated by postulating an isotropic state at $x=\infty $ ($P(\infty,t)=0$) and a polar state at
$x=-\infty $ ($P(-\infty,t)=1$). This gives us the ratio of the density in in a polarized region to
the density  an isotropic region  as
\begin{equation}
\frac{\rho _{pol}}{\rho _{iso}}=\frac{U}{U-v_{0}P} \;.
\end{equation}
The density contrast is infinitely sharp when the front
propagates at a speed $v_{0}P$ commensurate with the degree of ordering in
the stripes. Diffusion, neglected here, will smooth the density crossover between the two regions. This is in agreement with our numerical results that indicated that the contrast between the two regions is insensitive to the parameters and is indeed diffusion limited.

Writing  the density from Eq.~(\ref{densSimple}) as $\rho =1/(U-v_{0}P)$, and
substituting in the order parameter equation,  we can formally integrate
Eq.~(\ref{polSimple})  to obtain a solution of the form
\begin{equation}
x=-\frac{\Lambda }{\left( P-1\right) }+\frac{\Lambda a_{4}}{\zeta _{R}}\ln %
\left[ a_{4}(P-1)^{2}\left( 1+P\right) \right] +\frac{v_{0}}{2a_{2}}\ln (P),
\label{eq:front}
\end{equation}
where for simplicity we have assumed $U\sim v_{0}$ and we have introduced a dimensionless
friction constant, $\zeta _{R}=\left( a_{2}+a_{4}\right) $  and a dimensionless length scale, $\Lambda =\frac{\lambda _{3}}{v_{0}\zeta
_{R}}$. This formal
solution cannot be inverted analytically. A plot is shown in Fig.~\ref{fig:front} 
and clearly displays that the solution represents a propagating domain boundary of effective thickness  $\Lambda $ between a state with $P=1$ and a state with $%
P=0$.  In
physical units, the  length scale controlling the crossover between isotropic and polarized states, hence the sharpness of the stripes, is given by $\left( 3\pi
v_{0}/\ell D_{r}\right) $. This length  is essentially the distance
traveled by a self propelled particle in a  rotational diffusion time.
In other words, stripe formation is controlled by the formation of domain boundaries in the polarization, and the fact that
the density is slaved to the polarization and hence leads to mass fluxes that delineate the two regions in the striped phase. Finally, it is apparent from Eqs.~(\ref{densSimple}) and (\ref{polSimple}) that there is no
propagating front solution if we turn off the couplings between density
and polarization.
\begin{figure}[tbp]
\begin{center}
{\includegraphics[height=9.0cm, width=9.0cm,angle=270]{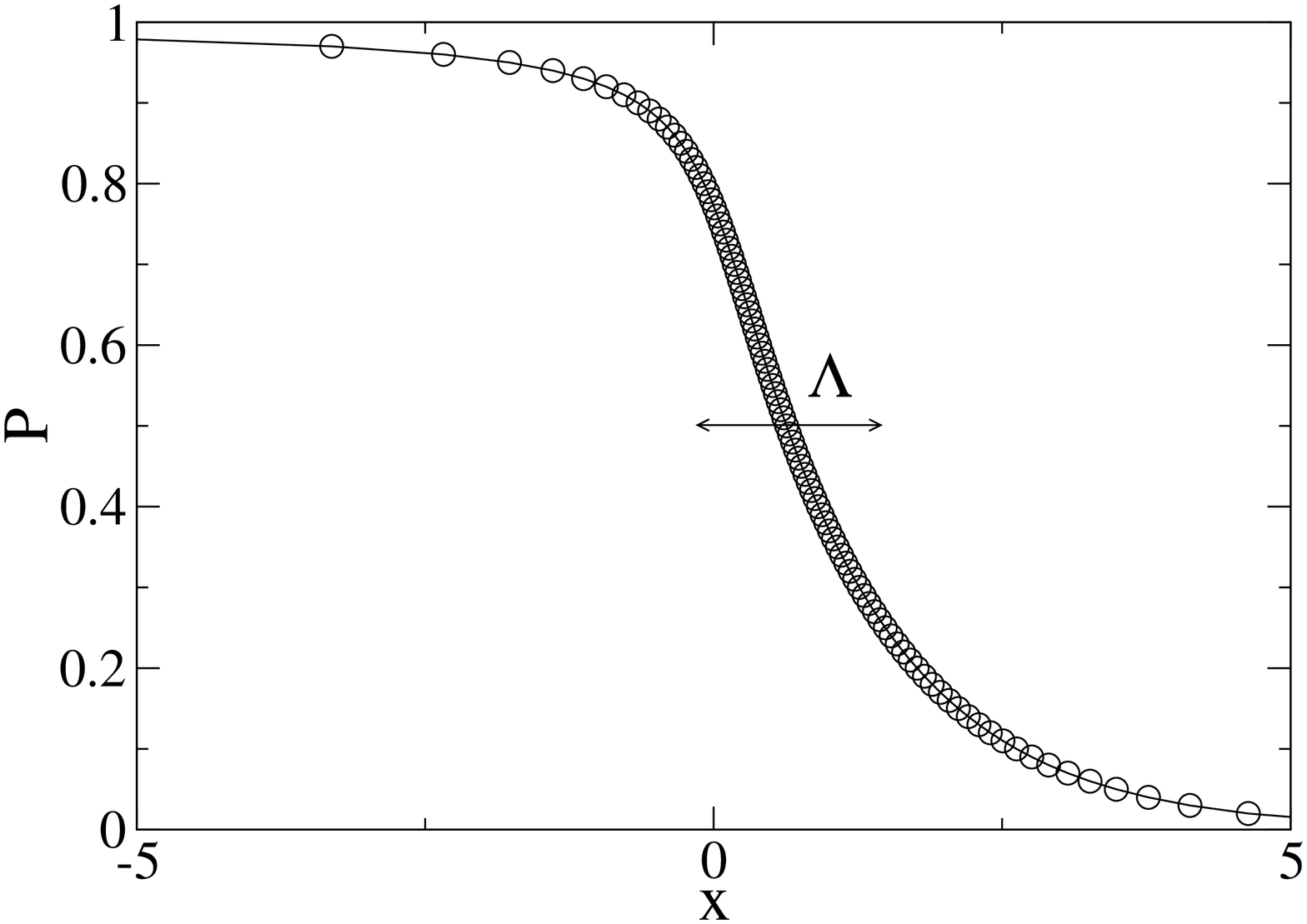}}
\end{center}
\caption{Plot of polarization as a function of $x$ as obtained by inverting Eq. \ref{eq:front} for $\rho_0 = 0.7$ and $v_0 = 1.0$. The solution represents a propagating domain boundary between a state with $P=1$ and a
state with $P=0$. For these parameters $\Lambda \sim 1.57$.}
\label{fig:front}
\end{figure}

\section{Summary}

In this work we have considered a continuum description of a collection of self propelled particles moving in a passive medium. Their dynamics on large length and time scales is governed by hydrodynamic equations for the density and the polarization field. The crucial physics in these systems that distinguishes them from conventional liquid crystalline systems is the dual role of the polarization field as i) a physical velocity that leads to mass convection and hence couples orientational fluctuation to density fluctuations and ii) an order parameter associated with a spontaneously broken continuous symmetry. This duality leads to the remarkable phenomenon of long range ordering identified in \cite{tonertuPRL95}. Here we show that this same physics destabilizes the homogeneous ordered state above a critical value of self propulsion speed and allows the nonlinear equations to admit a propagating front solution that yields the striped phase identified  numerically. Further, the coupling of orientational fluctuations to density fluctuations gives rise to anomalous fluctuations even in the regime where the ordered state is stable and leads to nontrivial coarsening dynamics that is different from the dynamics of both the equilibrium 2D X-Y model and that of active nematics.

The two phases observed here, namely the striped phase and the fluctuating flocking phase, have been identified earlier in the context of numerical studies of the Vicsek model. Our work identifies the origin of these phenomena in the model independent framework of the dynamics of conserved quantities and broken symmetry variable. It has been shown in different systems of this class that pattern formation phenomena might be crucially related to biological functionality \cite{MyxoRipple, ReviewSwarming}. This work would facilitate the application of theoretical tools, such as the amplitude equations and pattern selection analysis that are well developed in the context of chemical reacting systems to collections of self propelled particles.

\begin{acknowledgements}
This work was supported by NSF grants DMR-075105 and DMR-0806511. We thank S. Ramaswamy  for helpful discussions.
\end{acknowledgements}

\appendix
\section{Linearized Equations}

Here we analyze in more detail the hydrodynamic equations  linearized about the homogeneous polar state
given in Eq.~(\ref{eqnFormal}) and (\ref{fullLinear}) in the main text and present a better approximation for the discussion of the splay instability.

The hydrodynamic modes deep in the ordered phase were discussed in Section III  by entirely neglecting magnitude fluctuations $\delta P$ that decay on microscopic time scales. A better approximation consists of  neglecting the rate of change of $\delta P$ in Eq.
~\eqref{polmaglin} and solve for $\delta P$ in terms of fluctuations in
density and director to lowest order in gradients, with the result
\begin{eqnarray}
\delta \tilde{P} &=&\frac{1}{2a_{20}}\Big\{\left[ 2a_{20}\alpha P_{0} +\imath k_\Vert\left( \frac{v_{0}}{2}-
\tilde{\lambda}\rho _{0}p_{0}^{2}\right)\right] \frac{
\delta \tilde{\rho} }{\rho _{0}}  \notag \\
&&+ \imath k_\perp\rho _{0}\lambda _{2}P_{0}^{2}
\delta \tilde{p}_{\bot }\Big\}
\end{eqnarray}
We then use this expression to eliminate $\delta\tilde{P}$ from from Eqs.~\eqref{eqnFormal} for density and director fluctuations. The eigenvalues of the resulting two coupled equations are given
\begin{equation}
s_{\pm }=\frac{1}{2}(c_{11}+c_{22})\pm \frac{1}{2}\sqrt{
(c_{11}-c_{22})^{2}+4c_{12}c_{21}}  \label{modes_formal}
\end{equation}
where
\begin{widetext}
\begin{eqnarray}
c_{11} =ik_\Vert v_{0}P_{0}\left( 1+\alpha \right)
-k_{\Vert}^{2}\left[D
-\frac{1}{2a_{20}}\left( \frac{v_{0}^{2}}{2}-\tilde{\lambda}v_{0}\rho
_{0}P_{0}^{2}\right) \right]
\end{eqnarray}
\begin{eqnarray}
c_{12} =ik_\perp v_{0}\rho _{0}P_{0}
-k_\Vert k_\perp \rho _{0}^2v_{0}^{2}P_0^2\frac{\lambda _{2}}{2a_{20}}
\end{eqnarray}
\begin{eqnarray}
c_{21} =ik_\perp \left[ {\frac{v_{0}}{2P_{0}}}-\rho _{0}\lambda _{3} P_{0}\left( 1+\alpha\right) \right]
-k_\Vert k_\perp\left[\left( D_{s}-D_{b}\right)
-\frac{\lambda _{3}\rho _{0}}{2a_{20}}\left( \frac{v_{0}}{2}-\overline{
\lambda}\rho _{0}P_{0}^{2}\right) \right]
\end{eqnarray}
\begin{eqnarray}
c_{22} =-ik_\Vert \lambda _{1}\rho _{0}P_{0}-\left[D_{b}k_{\Vert}^{2}
+\left( D_{s}-\frac{\lambda _{3}\lambda _{2}\rho _{0}^{2}}{2a_{20}}
P_{0}^{2} \right)k_{\perp}^{2}\right]
\end{eqnarray}
\end{widetext}
Again the modes decouple when ${\bf k}=k_\Vert {\bf \hat{p}}_0$  lies along the direction of broken symmetry. The two modes governing the dynamics of density and director fluctuations are then given by
\begin{subequations}
\begin{gather}
s_\rho^L=ikv_{0}P_{0} \left( 1+\alpha \right) -\left[D-\frac{1}{4a_{20}}\left( v_{0}^2+2\overline{\lambda}v_{0}\rho
_{0}P_{0}^{2}\right) \right] k^{2}\;,\label{s-rho}\\
s_p^L=-ik\lambda _{1}\rho _{0}P_{0}-D_{b}k^{2}\;.
\end{gather}
\end{subequations}
Director fluctuations are stable and their decay rate is controlled by the bend diffusion constant. Since $a_{20}<0$ in the ordered state, one can define an effective diffusion constant in Eq.~(\ref{s-rho})
as $D_{eff}^L=D+(v_0^2+\overline{\lambda}v_{0}\rho
_{0}P_{0}^{2})/(4|a_{20}|)$. The first correction to $D$ in this expression, proportional to $v_0^2$,  always enhances  diffusion and arises from the fact that self
propelled particles perform a persistent random walk \cite{aparna}. The second correction can lead to an instability if $\overline{\lambda}<0$. The parameters $\lambda_i$ are microscopic quantities and their values are model dependent.
As discussed in the main text, if we think of the  polarization an an equilibrium  order parameter, then $ \overline{\lambda}=0$. If in contrast we think of the polarization as a  physical velocity in a Galilean invariant system, then  $\overline{\lambda}>0$. In both these cases the density fluctuations  relax diffusively for all values of the parameters.
For the self-propelled hard rod model discussed in Ref.~\cite{aparna}
$\overline{\lambda}>0$ (see also Table~\ref{table1}). In this case the convective terms proportional to $\lambda_i$ further enhance the effective diffusion
constant and the homogeneous state is stable. Note that a value $\overline{\lambda}>0$ is also obtained in the microscopic Boltzmann equation model studied in \cite{bdg}. If, however,
the microscopic model allows for higher order chemical and biological
processes that can lead, for example, to a reversal of the direction of motion
of an individual unit due to interactions with other units, then $\overline{
\lambda}$ can be negative and drive a longitudinal instability \cite{cates}.

Next, we consider wavevectors ${\bf k}$ transverse to the direction $\hat{\bf p}_0$ of broken symmetry, i.e., $k=k_\perp$. In this case the equations for density and director fluctuations are coupled and the two hydrodynamic modes are given by
\begin{widetext}
\begin{equation}
s_{\pm }^{T}=-\frac{1}{2}\left( D+\overline{D}_{s}\right) k^{2}\pm
\frac{1}{2} \Big\{\left( D-\overline{D}_{s}\right)
^{2}k^{4}-2k^{2}v_{0}\rho _{0}\left[
v_{0}-2\rho _{0}P^{2}_{0}\lambda _{3}\left( 1+
\alpha\right) \right] \Big\}^{1/2},
\label{splaymodesA}
\end{equation}
\end{widetext}
where $\overline{D}_{s}=D_{s}+\rho _{0}^2P_0^2\lambda _{2}\lambda _{3}/(2|a_{20}|)$.
The mode $s_{+}^{T}$ can become positive, yielding an instability, for $k<k_c$,
with
\begin{equation}
k_c=\sqrt{v_0\left[2\rho_0P_{0}^{2}\lambda_3(1+\alpha)-v_0\right]/(D\overline{D}_s)},
\label{kcA}
\end{equation}
provided
\begin{equation}
2\rho_0P_{0}\lambda_3[1+\alpha]>v_0.
\label{instA}
\end{equation}
Using the parameter values obtained for the model of self-propelled hard rods discussed in \cite{aparna},
where the nonlinear terms in the polarization equation arise from
momentum-conserving collisions between the self-propelled rods,
and summarized in Table~\ref{table1}, we obtain
\begin{equation}
v^{S}_c=\left[2\pi\rho _{0}P_{0}^{2}\left( 1+\alpha \right)\right]^{-1}.
\label{eqvcA}
\end{equation}
The critical line $v^{S}_c(\rho_0)$ given in Eq.~(\ref{eqvcA}) is plotted in Fig.~\ref{fig:phasedia}. As obtained in the main text , the instability line vanishes as $v^{S}_c\sim1/\rho_0$ at large density. However, near the transition incorporating overdamped magnitude fluctuations  regularizes the behavior yielding a finite value for $v^{S}_c(\rho_c)$.  Finally, for a wavevector ${\bf k}$ at an angle $\theta$ to direction $\hat{\bf p}_0$ of broken symmetry,
the splay instability occurs for a range of angles $\theta_m\leq\theta\leq\pi/2$, where $\theta=\pi /2$ corresponds to ${\bf k}$ normal to $\hat{\bf p}_0$. The growth rate of the unstable mode is, however, always largest for $\theta=\pi /2$, when director deformations are pure splay.

\section{Wavevector of fastest growing modes}
In this appendix we identify and charachterize the fastest growing mode associated with the two linear instabilities identified in the main text. To identify the wavevector of the fastest growing mode for the longitudinal instability discussed in Section III, we evaluate the real part of the dispersion relation of this mode to order $k^4$, with the result
\begin{equation}
Re[s_-^L(k)]=-D_{eff}k^2-D_4k^4+{\cal O}(k^6)\;,
\end{equation}
where $D_{eff}$ is given in Eq.~\eqref{deff} and
\begin{widetext}
\begin{eqnarray}
D_{4} &=&\frac{1}{32\left| a_{20}\right| ^{3}}A\left( A-4\alpha
P_{0}v_{0}\right) (D-D_{s})  \nonumber \\
&&+\frac{1}{32\left| a_{20}\right| ^{2}}\left[ \left( \frac{1}{8\left|
a_{20}\right| }\left( A-4\alpha P_{0}v_{0}\right) ^{2}+\frac{1}{\left|
a_{20}\right| }v_{0}^{2}\right) ^{2}+4\left( D-D_{s}\right) \left( \frac{1}{%
8\left| a_{20}\right| }\left( A-4\alpha P_{0}v_{0}\right) ^{2}+\frac{1}{%
\left| a_{20}\right| }v_{0}^{2}\right) \right]   \nonumber \\
&&+\frac{3}{16\left( 8\left| a_{20}\right| ^{3}\right) }A^{2}\left( -2\left(
D-D_{s}\right) -\frac{1}{8\left| a_{20}\right| }\left( A-2\alpha
P_{0}v_{0}\right) ^{2}-\frac{1}{\left| a_{20}\right| }v_{0}^{2}\right) -%
\frac{5}{128\left( 2\left| a_{20}\right| \right) ^{8}}A^{4}
\end{eqnarray}
\end{widetext}
where
\begin{equation}
A=2\left[v_0(1-2\alpha)+\overline{\lambda }\rho_0\right]P_0
\end{equation}
\begin{figure}[tbp]
\begin{center}
{\includegraphics[height=7.0cm, width=7.0cm]{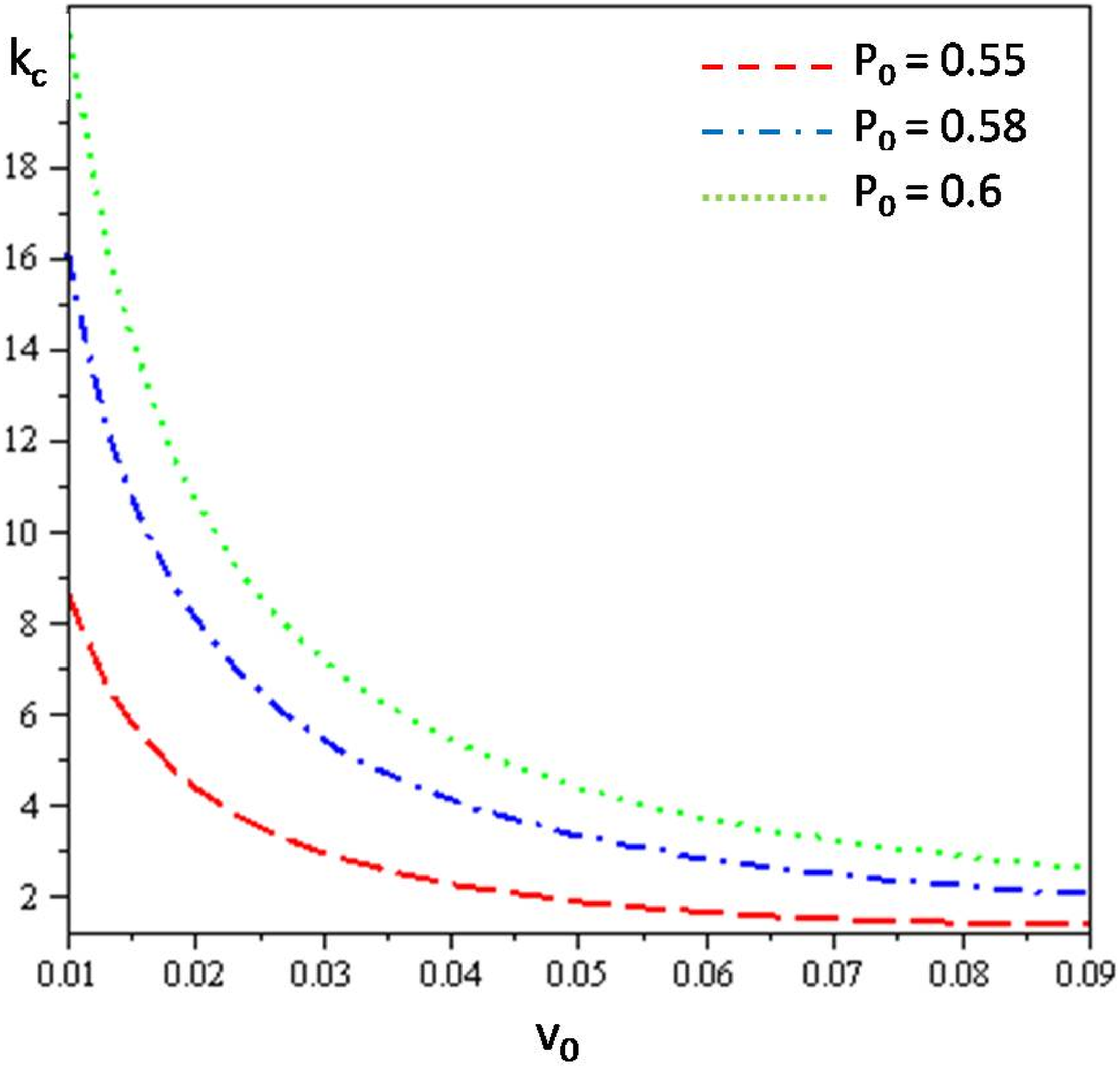}}
\end{center}
\caption{(color online) The maximum growth wavevector of the longitudinal instability as a function of the self-propulsion speed $v_0$ for different densities.}
\label{fig:kc}
\end{figure}
\begin{figure}[tbp]
\begin{center}
{\includegraphics[height=7.0cm, width=7.0cm]{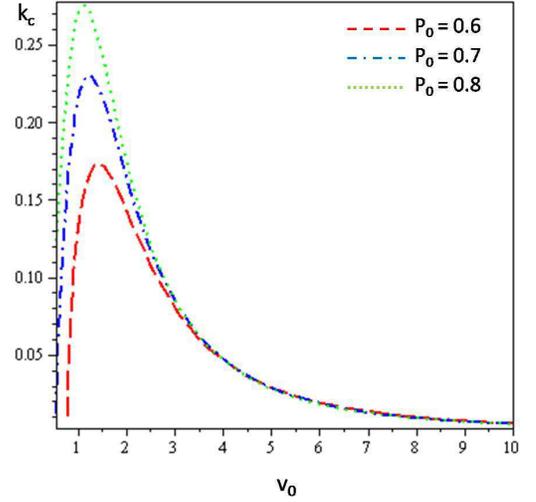}}
\end{center}
\caption{(color online) The maximum growth wavevector of the splay instability as a function of the self-propulsion speed $v_0$ for different densities.}
\label{fig:kc2}
\end{figure}
The growth rate of the unstable mode is then maximum at a wavevector $k_c$, given by
\begin{equation}
k_{c}^{L}=\sqrt{-D_{eff}/2D_{4}}
\end{equation}
Fig.~(\ref{fig:kc}) shows a plot of the maximum growth wavevector as a function of the self propulsion speed for various values of the mean density $\rho_0$. The critical length scale $k_c^{-1}$ decreases with increasing density $\rho_0$, in agreement with what observed numerically for the width of the stripes. On the other hand, $k_c^{-1}$ increases with increasing $v_0$decreases as the self propulsion speed increases, implying that the width of the stripe should increase with increasing SP speed, while the stripes width exhibits the opposite behavior.   This indicates that the length scale selected by the nonlinear pattern is not simply related to the wavevector of the most unstable mode associated with the linear instability.

Next, proceeding as above, we can find the fastest growing mode associated with the splay instability. This is of the form
\begin{equation}
k_c\sim \frac{v_{0}}{(
\overline{D}_{s}+D)}~\sqrt{2\left( \frac{v_{0}}{v_{c}}-1\right) }\;.
\end{equation}
This critical wavevector is shown as a function of self-propulsion speed $v_0>v^{S}_c(\rho)$ for different values of $\rho_0$ in Fig.~(\ref{fig:kc2}). In this case $k_c$ is a non-monotonic function of $v_0$. But, it increases with both $v_0$ and $\rho_0$ for the range of densities and self propulsion speeds probed by the numerical analysis.


\end{document}